\documentclass[sigconf]{acmart}

\usepackage{multirow}
\usepackage{url}
\usepackage{makecell}
\usepackage[table]{xcolor}

\AtBeginDocument{%
  }

\copyrightyear{2026}
\acmYear{2026}
\setcopyright{cc}
\setcctype{by-nc-nd}
\acmConference[CHI '26]{Proceedings of the 2026 CHI Conference on Human Factors in Computing Systems}{April 13--17, 2026}{Barcelona, Spain}
\acmBooktitle{Proceedings of the 2026 CHI Conference on Human Factors in Computing Systems (CHI '26), April 13--17, 2026, Barcelona, Spain}
\acmPrice{}
\acmDOI{10.1145/3772318.3790654}
\acmISBN{979-8-4007-2278-3/2026/04}

\begin{document}

% \title[Conversing with LLM-based AI Homogenizes Our Self-Views]{Conversing with Large Language Model Based AI Homogenizes Our Self-Views}
% \title[Self-view Alignment with AI]{Self-view Alignment with AI: Large Language Model Expressed Personality Traits Shape and Homogenize Human Self-views}
\title{AI-exhibited Personality Traits Can Shape Human Self-concept through Conversations}


%
\author{Jingshu Li}
\orcid{0009-0006-1576-8487}
\affiliation{%
  \department{Computer Science}
  \institution{National University of Singapore}
  \city{Singapore}
  \country{Singapore}
}
\email{jingshu@u.nus.edu}

\author{Tianqi Song}
\orcid{0000-0001-6902-5503}
\affiliation{%
  \institution{National University of Singapore}
  \city{Singapore}
  \country{Singapore}
}
\email{tianqi_song@u.nus.edu}

\author{Nattapat Boonprakong}
\orcid{0000-0002-0735-4536}
\affiliation{%
  \department{School of Computing}
  \institution{National University of Singapore}
  \city{Singapore}
  \country{Singapore}
}
\email{nattapat.boonprakong@nus.edu.sg}

\author{Zicheng Zhu}
\orcid{0000-0002-4332-2515}
\affiliation{%
  \department{School of Computing}
  \institution{National University of Singapore}
  \city{Singapore}
  \country{Singapore}
}
\email{zicheng@u.nus.edu}

\author{Yitian Yang}
\orcid{0009-0000-7530-2116}
\affiliation{%
  \department{School of Computing}
  \institution{National University of Singapore}
  \city{Singapore}
  \country{Singapore}
}
\email{yang.yitian@u.nus.edu}

\author{Yi-Chieh Lee}
\orcid{0000-0002-5484-6066}
\affiliation{%
  \institution{National University of Singapore}
  \city{Singapore}
  \country{Singapore}
}
\email{yclee@nus.edu.sg}

% \renewcommand{\shortauthors}{Li et al.}

% \newcommand{\rhl}[1]{\textcolor{blue}{#1}}
\newcommand{\rhl}[1]{{#1}}
\begin{abstract}

Recent Large Language Model (LLM) based AI can exhibit recognizable and measurable personality traits during conversations to improve user experience. However, as human understandings of their personality traits can be affected by their interaction partners' traits, a potential risk is that AI traits may shape and bias users' self-concept of their own traits. To explore the possibility, we conducted a randomized behavioral experiment. Our results indicate that after conversations about personal topics with an LLM-based AI chatbot \rhl{using GPT-4o default personality traits, users' self-concepts aligned with the AI's measured personality traits.} The longer the conversation, the greater the alignment. This alignment led to increased homogeneity in self-concepts among users. We also observed that the degree of self-concept alignment was positively associated with users' conversation enjoyment. Our findings uncover how AI personality traits can shape users' self-concepts through human-AI conversation, highlighting both risks and opportunities. We provide important design implications for developing more responsible and ethical AI systems.
\end{abstract}






\begin{teaserfigure}
\centering 
\includegraphics[width=\textwidth]{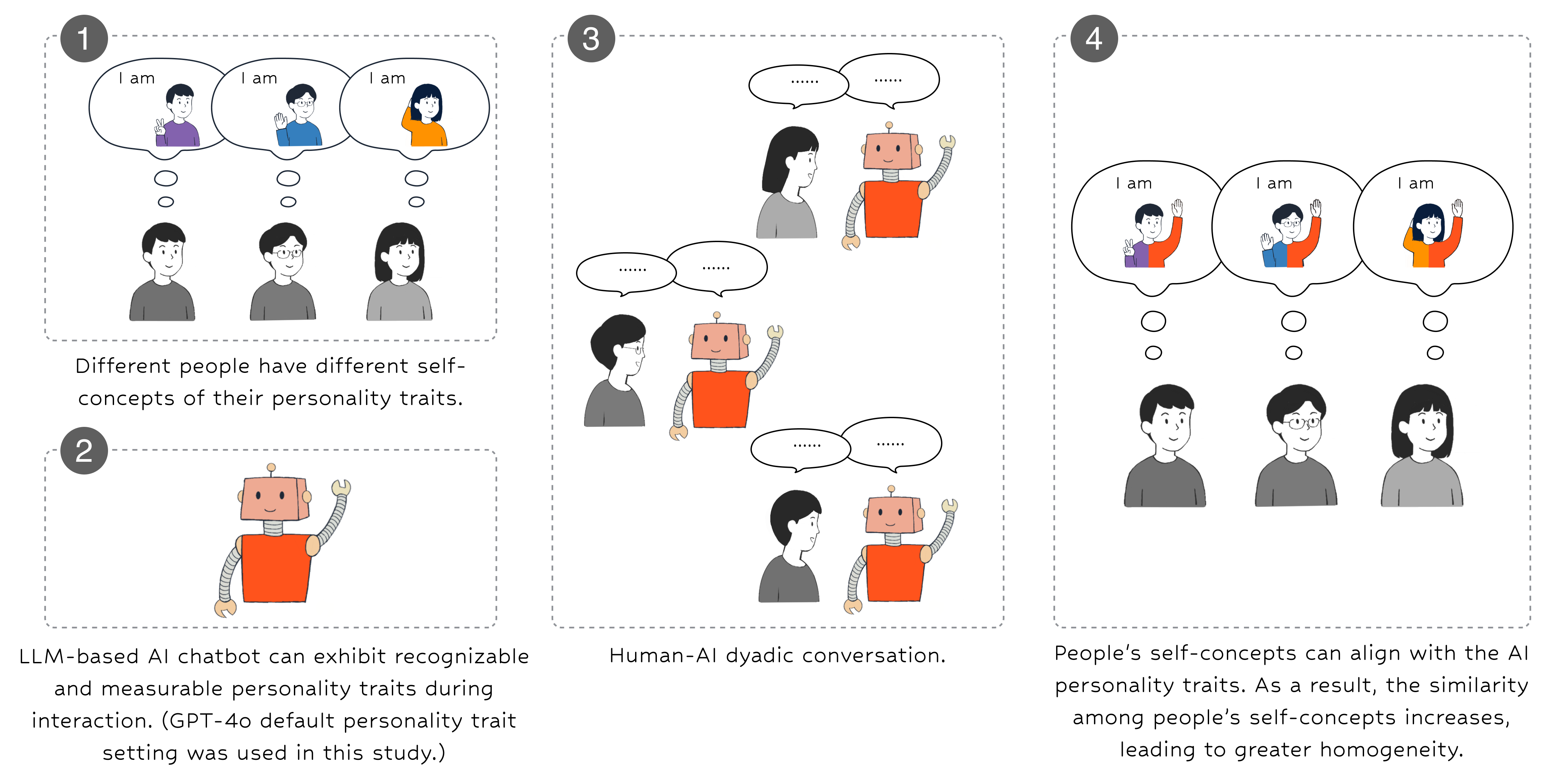} 
\caption{Our study shows how LLM-based AI chatbot's personality traits can shape human self-concept through conversation. Different individuals have different self-concepts about their own personality traits, and current Large Language Model (LLM) based AI systems are also capable of simulating and exhibiting \rhl{recognizable and measurable} personality traits. \rhl{In this study, GPT-4o default personality trait setting was used, and we measured AI traits by scales completed by the AI chatbot.} Through conversations with the same AI, each individual’s self-concept becomes \rhl{aligned with the AI's measured personality traits}, leading to increased homogeneity of self-concepts across individuals.}
\label{fig: teaser}
\Description{Illustration of how AI personality traits shape human self-concept through conversation. The figure on the left depicts that different individuals have different self-concepts about personality traits, and current LLM-based AI systems are also capable of exhibiting personality traits. The middle figure represents the human-AI conversation. The figure on the right shows that through conversations with the same AI, each individual’s self-concept becomes aligned with the personality traits exhibited by the AI, leading to increased homogeneity of self-concepts across individuals.}
\end{teaserfigure}


\maketitle

\section{Introduction}
Humans can perceive their own personality traits, i.e., patterns of thoughts, feelings, and behaviors that distinguish individuals from one another~\cite{matthews2003personality, markus1987dynamic}.
For example, someone who often feels energized by social interactions may recognize extraversion as part of their personality traits.
Through this process, they construct \textit{self-concept} of these traits~\cite{markus1987dynamic, hampson2019construction}. One's self-concept is highly subjective; it embodies the answer to the question ``\textit{Who am I?}''~\cite{myers2009social}. Psychologists suggest that self-concept significantly shapes the way individuals navigate decisions, goals, and social interactions in daily life~\cite{gallagher2011oxford, swann2007people}. 
For example, individuals with a self-concept of high extroversion are more likely to set career goals centered on social interaction (e.g., sales) and to actively engage in social contexts~\cite{wille2012transactional}. Similarly, individuals whose self-concept reflects high conscientiousness tend to establish more organized, long-term goals and persist in their academic or professional pursuits~\cite{roberts2000broad}.

Self-concept constantly changes: individuals develop their self-concepts through previous encounters with the world~\cite{markus1987dynamic}. More importantly, social interactions can shape one's self-concept~\cite{markus1987dynamic,hampson2019construction}.
Previous research has indicated that individuals' self-concepts tended to shift during conversations~\cite{john2010handbook, corr2020cambridge}. 
The personality traits of conversation partners can influence the way individuals view themselves~\cite{welker2024self, van2020codevelopment}. For example, \citet{welker2024self} has shown that individuals \textit{align} self-concepts with their conversation partners, even in a one-time conversation. 
Different topics of conversation can also determine the extent of change in self-concept~\cite{tice1992self, chen2006relational}. 
Especially, in conversations where self-disclosure and self-reflection likely occur, such as conversations about personal topics, individuals' self-concepts tend to change more~\cite{welker2024self, sun2020well, mehl2010eavesdropping}.

Today, social interactions can take the form of human-AI interaction. People now engage with LLM-based AI chatbots to share personal experiences and seek companionship as well as emotional support \cite{chaturvedi2023social, zhang2025dark, meng2021emotional}. AI chatbots have been increasingly equipped with human-like personality traits (e.g., Instagram AI Studio\footnote{https://aistudio.instagram.com/} and Character.AI\footnote{https://character.ai/}) through prompts or fine-tuning~\cite{jiang2023personallm, wu2025incorporating, shao2023character}. 
Recent HCI studies have pointed out that such design choices present several benefits: they can increase the perceived humanness of AI chatbots~\cite{li2025exploring, jones2025artificial}; encourage continued user engagement~\cite{wu2025incorporating}; improve emotional resonance and sense of safety~\cite{jones2025artificial}; and foster long-term harmonious human-AI relationships~\cite{pataranutaporn2021ai, li2025exploring}.

\rhl{Although whether AI possesses genuine personality traits remains a topic of debate~\cite{placani2024anthropomorphism, rescorla2015computational}, prior work indicates that the impact of conversation partner's personality traits on one's self-concept depends not on the partner’s genuine personality but on perceivable cues such as linguistic style and expression patterns~\cite{welker2024self, van2020codevelopment, markus1987dynamic}. LLM-based systems display stable and consistent language patterns from which humans can reliably infer configured traits~\cite{jiang2023personallm}. Such simulated personality traits, conveyed through stable cues, may evoke interpersonal-like psychological processes that shape humans’ self-concepts, yet alignment is not guaranteed because humans may treat AI as a tool or as lacking genuine agency, which can attenuate the influence of these cues~\cite{li2025exploring, momen2024social, ahn2024role}. Nevertheless, recent HCI studies show that AI can exert social influence across interaction contexts, aligning humans’ behavior or cognition with the AI~\cite{li2025we, shen2024towards}. Building on this literature on human-AI alignment~\cite{li2025we, shen2024towards} and on individuals’ shifts in self-concept~\cite{welker2024self, van2020codevelopment, markus1987dynamic}, we investigate how interacting with AI chatbots exhibiting human-like personality traits may potentially influence human self-concepts. In other words, users may align their self-concepts with the AI personality traits.}

People's self-concept alignment with AI may carry notable psychological and social impacts.
On the one hand, it poses risks both at the individual and group levels.
At the individual level, self-concept alignment with AI can hinder accurate conceptualization of self. Consequently, inaccurate self-concepts can undermine an individual's well-being and interpersonal relationships~\cite{colvin1994positive, kwan2008conceptualizing, corr2020cambridge}.
At the group level, influenced by social and cultural factors, self-concept varies across individuals~\cite{gallagher2011oxford,hampson2019construction, matthews2003personality}. 
However, self-concept alignment could occur collectively, especially when different users interact with the same conversational AI model~\cite{chaturvedi2023social, zhang2025dark, meng2021emotional}. Subsequently, this could reduce the diversity of self-concepts shaped by different social and cultural backgrounds~\cite{gallagher2011oxford, hampson2019construction}, and further impair inclusiveness and collective intelligence~\cite{page2008difference}.
On the other hand, self-concept alignment may also have potential benefits. Alignment in self-concepts could increase the experience of shared inner states (i.e., shared reality experience~\cite{echterhoff2009shared}) during human-AI conversation. The increased shared reality experience can further elicit positive emotions and enhance the conversation experience~\cite{wagner2015beautiful,reece2023candor}.
Therefore, understanding human-AI self-concept alignment would reveal deeper psychological and social impacts of AI, highlight potential risks in current AI applications, and support the development of more responsible and ethical AI design. However, limited research has explored the process of human self-concept alignment with LLM-based AI.

To examine the influence of AI personality traits on users’ formation of self-concept, we conducted an online randomized behavioral experiment with a mixed factorial design. 
During the experiment, participants engaged in a conversation with an LLM-based AI chatbot \rhl{operating under GPT-4o's default personality trait settings. The AI chatbot’s personality traits were measured using a personality scale completed by the AI itself.}
The participants' self-concept (pre- \textit{vs.} post-conversation) was the within-subject factor.
Meanwhile, human-AI conversations cover a wide range of topics~\cite{fang2025ai}.
To explore the effect of different conversation topics, inspired by previous research~\cite{welker2024self, fang2025ai, jeon2025letters}, we included the conversation topic (personal topics \textit{vs.} non-personal topics) as a between-subject factor and randomly assigned participants to two conditions accordingly.

Our results indicate that, in conversations about personal topics, participants' self-concepts aligned \rhl{with the AI chatbot's measured personality traits}--a process we call \textit{self-concept alignment with AI}. 
The degree of self-concept alignment is positively correlated with the conversation length.
Specifically, we found that self-concept alignment with AI made participants' self-concepts more similar, increasing the homogeneity of self-concepts. Figure~\ref{fig: teaser} gives an overview of our findings.
The alignment also showed a benefit: the degree of alignment between participants’ self-concepts and AI personality traits was positively associated with conversation enjoyment. Further analysis revealed that this effect was serially mediated by the accuracy of participants’ perception of AI personality traits and participants’ shared reality experience during the conversation.

In sum, our study makes the following contributions to the HCI community:
\begin{itemize}
    \item We demonstrate that human self-concepts can align with \rhl{specific AI's measured personality traits} during human-AI conversation. It suggests that, in addition to the influence of social interactions with other humans, personality traits exhibited by the AI conversation partner can influence human self-concepts. We further reveal and discuss the double-edged role of human-AI self-concept alignment.
    \item \rhl{We identify the effect of conversation topic on the self-concept alignment with AI and the correlation between conversation length and the alignment. We} further explain the mechanism through which self-concept alignment influences conversation enjoyment. Our findings inform the understanding of self-concept changes in human-AI interaction. 
    \item We present important design implications for the risks and opportunities of human-AI self-concept alignment. We provide new insights for developing more responsible and ethical conversational AI systems in the future.
\end{itemize}

\section{Related Work}
This study is grounded in previous research in HCI, AI, and psychology. First, we review research on the personality traits of LLM-based AI, which provides an important technical foundation for this study. Second, we discuss existing HCI research that investigates the alignment of human behavior and cognition with AI, which inspired the design of this study. Finally, we draw on literature from psychology and HCI concerning self-concept and its dynamics, which provides the theoretical foundation of our study. We discuss the important implications of self-concept and the factors that shape self-concept.

\subsection{Personality Traits in LLM-based AI}
In the context of LLM, personality traits refer to human-like distinguishable patterns of thoughts, feelings, and behaviors simulated by the model~\cite{pan2023llms}. These traits are influenced by training data and prompts, and can affect the model’s expressed behavior and opinions during interaction~\cite{wen2024self}. As LLMs exhibit increasingly human-like qualities in text-based interactions, researchers in AI, HCI, and psychology have shown a growing interest in LLM personality traits~\cite{jiang2023personallm,serapio2023personality,pellert2024ai}.

One key question researchers are looking at is how to measure the personality traits of LLM~\cite{serapio2023personality}. The most direct method is to use personality questionnaires designed to assess traits. Many psychological scales have been applied to measure LLM personality traits~\cite{dong2025humanizing}, such as the Big Five Personality Inventory~\cite{jiang2023evaluating,john1991big}, MBTI~\cite{jiang2023personallm,briggs1976myers}, SD-3~\cite{li2022evaluating,jones2014introducing}, and HEXACO~\cite{ashton2009hexaco,miotto2022gpt}. Some studies have also developed trait assessment tools specifically for LLMs, such as the Machine Personality Inventory~\cite{jiang2023evaluating} and TRAIT~\cite{lee2024llms}. Researchers typically prompted LLMs to complete these questionnaires (often multiple times), and then extracted responses using methods such as regular expression analysis~\cite{jiang2023evaluating}, token probability analysis~\cite{pan2023llms}, or constrained output formats~\cite{la2025open}.
Beyond directly utilizing questionnaire items, some studies employed text analysis to infer LLM personality traits~\cite{karra2022estimating}. For example, \citet{karra2022estimating} used an end-to-end zero-shot classifier to categorize LLM responses according to the Big Five dimensions. Other approaches include using Linguistic Inquiry and Word Count (LIWC) or human raters to analyze the responses generated by LLMs to personality-related questions~\cite{jiang2023evaluating,frisch2024llm}.

Recent studies have also focused on the reliability and validity of these measurements~\cite{serapio2023personality,ye2025large}. From the perspective of psychometrics, researchers have examined personality trait measurements for LLMs~\cite{ye2025large}. For example, some studies adopted Cronbach's alpha and intraclass correlation coefficients to quantify the reliability of trait assessments for LLM~\cite{zheng2025lmlpa,huang2023revisiting,ceron2024beyond}. Recent findings suggested that LLMs can yield reliable results in personality assessments~\cite{huang2023revisiting,serapio2023personality}, and the language generated by LLMs often reflects consistent patterns associated with their measured traits~\cite{jiang2023personallm}.

At the same time, studies have explored how human users perceive AI personality traits~\cite{wen2024self}. For instance, \citet{jiang2023personallm} found that users could accurately perceive certain personality traits (e.g., extraversion and agreeableness) exhibited by GPT-3.5 and GPT-4 \cite{jiang2023personallm}.
This is not surprising, as language can convey personality traits~\cite{gosling2003very,mairesse2007using,pennebaker2003psychological}, and people can identify one’s personality traits from short language samples~\cite{biesanz2007you}.
However, other studies suggested that users' perceptions of a chatbot's traits may be influenced by their personal, subjective preferences~\cite{ruane2021user}.

\subsection{People Align with AI during Their Interaction with AI}

Research in HCI has suggested that humans actively align their behavior and cognition with AI during human-AI interaction, that is, by adopting behaviors similar to AI or maintaining the same cognitive orientation as AI~\cite{li2025we, shen2024towards}.
\citet{shen2023effects} showed that in a voice-based picture naming and matching task, participants' word choices aligned with those of their conversation partners. Moreover, people were more likely to align with machine partners than with human ones. The authors suggested that this may be because users consider the machine's perspective and make greater adjustments to accommodate its limited communicative abilities in order to maintain the interaction.
Studies have also shown that, to maintain effective communication, human-AI alignment can also occur at the conceptual level~\cite{cirillo2022conceptual,van2012conceptual,zhang2025align}. For example, \citet{cirillo2022conceptual} found that the conceptual level at which humans select words to describe objects (e.g., using a specific name vs. a categorical label) tends to converge with that of AI.
\citet{agarwal2025ai} showed that when using AI as a writing assistant, the content and style of human creators align with AI, which increases homogeneity in human group-level creation.

In addition, human-AI alignment can result from AI exerting social influence on users~\cite{li2025we}. A recent study on multi-agent systems found that multiple AI agents can form a social group that influences users, leading them to conform and align their views with the group~\cite{song2024multi}. In the context of human-AI decision-making, \citet{li2025confidence} showed that users tended to align their confidence levels with those expressed by AI, even when they only observe AI confidence without directly collaborating with it. Further, researchers suggested that this alignment may partly result from unconscious social contagion~\cite{li2025we, li2025confidence}.

These human-AI alignment phenomena can be a double-edged sword~\cite{li2025we,shen2024towards}. On the one hand, some researchers argued that alignment improves the efficiency of the human-AI interaction~\cite{li2025we, zhang2025align}. For example, alignment in vocabulary and language can enhance communication efficiency~\cite{shen2023effects}. On the other hand, researchers have raised concerns that such alignment can bring risks to users at both the individual and group levels. For example, at the individual level, confidence alignment may impair individuals’ ability to calibrate their own confidence, potentially affecting human-AI collaboration outcomes~\cite{li2025confidence}.
Alignment in opinions can also threaten user autonomy. Given the risk of hallucination in LLM-based AI, such alignment could contribute to the spread of misinformation~\cite{song2024multi}. 
At the group level, recent studies and perspectives have raised concerns about AI contributing to homogenization among humans~\cite{kosmyna2025your, agarwal2025ai, anderson2024homogenization, qin2025timing}. For example, in AI-assisted writing, alignment of human writing content with AI may dilute cultural differences and cause cultural colonization~\cite{agarwal2025ai}.

\subsection{People's Self-concept Changes during Their Interaction with Human and AI}

\textit{Self-concept} refers to how people understand themselves, including people's beliefs about their physical appearance, personality traits, skills, abilities, etc.~\cite{gallagher2011oxford,hampson2019construction, matthews2003personality}. 
In this study, we focus on the self-concept of personality traits, i.e., how people view their own personality traits.
Self-concept can be formed through observation and assessment of self, as individuals infer their own traits and attitudes based on their behavior~\cite{andersen1984self}.
Self-concept has important implications for psychological, behavioral, and social outcomes~\cite{gallagher2011oxford}. Previous studies have found that people's self-concepts can influence their well-being and behavior in the real world~\cite{swann2007people, markus1987dynamic, kwan2008conceptualizing}.
For example, research has shown that a self-concept consistent with personality traits is necessary for healthy interpersonal relationships and well-being~\cite{kwan2008conceptualizing, corr2020cambridge}. 
Inconsistent self-concept, in contrast, can create conflict between behavior and the true self, leading to negative emotions such as frustration and disappointment~\cite{kim2011emotional, swann2007people}.
Positive self-concept, which refers to an individual's positive beliefs about themselves (e.g., a student believing that they are conscientious), can predict students’ academic performance~\cite{robbins2004psychosocial, kim2014effects}.
In contrast, negative self-concept has been found to predict aggression or antisocial behaviors~\cite{donnellan2005low}.

Self-concept is not static, and existing research suggests that self-concept is active, forceful, and capable of changing~\cite{markus1987dynamic}. 
Meaningful social interactions can influence self-concept~\cite{hampson2019construction, markus1987dynamic}. 
For example, conversation is a primary means by which humans maintain social bonds~\cite{dunbar1996grooming}, and self-concept can change during conversations~\cite{corr2020cambridge, john2010handbook}.
Previous studies have shown that in conversations with more self-disclosure and self-reflection, such as discussing personal topics, self-concept is more likely to change~\cite{schneider2024simulation, sun2020well, mehl2010eavesdropping}.

Furthermore, self-concepts tend to converge during dyadic social interactions~\cite{van2020codevelopment, welker2024self}.
This is in line with research on \textit{social influence}, which refers to the process by which the presence, words, behaviors, or attitudes of others affect an individual’s thoughts, emotions, or actions~\cite{cialdini2004social}.
Research has shown that a group of friends can influence each other over time and may come to see themselves as more similar in certain traits, such as extroversion~\cite{van2020codevelopment}.
\rhl{\citet{welker2024self} has further found a process called \textit{inter-self alignment}, indicating that even in short conversations, the self-concepts of both parties, that is, their understanding of their own personality traits, tend to converge. They also proposed that greater convergence in self-concept predicts higher conversational enjoyment: this relationship is mediated by the accuracy of individuals’ predictions about their partners’ personality traits~\cite{welker2024self}. 
This occurs because, when people engage in mental simulation and make inferences about others, they tend to draw upon their own self-concepts~\cite{waytz2011two, buckner2007self}. When self-concepts are aligned, predictions about others become more accurate, which increases perceived similarity~\cite{kenny1987accuracy, davis1996effect}. This, in turn, gives individuals the experience of having common inner states (such as feelings, beliefs, and evaluations) with others, known as \textit{shared reality}~\cite{echterhoff2009shared}. Shared reality experience has been shown to enhance positive emotions~\cite{wagner2015beautiful, reece2023candor}, thereby increasing conversational enjoyment.}

HCI research has explored changes in people's self-concept during their interaction with computing systems. 
Studies have shown that AI feedback can directly influence users’ self-concept~\cite{parsakia2023effect}. \citet{hung2025efficacy} found that AI-based positive feedback interventions -- designed with positive psychology principles -- can improve user positive self-concept. In contrast, negative AI feedback has the opposite effect. 
For instance, criticism from AI facial assessment tools can lead to high self-objectification and stronger negative self-concept of the user~\cite{agrawal2025psychological}, while negative AI feedback in the workplace can reduce positive self-concept of the user~\cite{li2025impact}. HCI studies have also used AI to create self-clones to promote self-reflection and change self-concept~\cite{lo2025d, jeon2025letters}. \citet{jeon2025letters} examined how interacting with an LLM simulating the user's future self influences the user's belief of their future self. \citet{zheng2025learning} designed AI-generated digital clones with the user's self-image to enhance positive self-concept during presentation.

\section{Hypotheses Development}
As AI (especially AI based on LLM) becomes more human-like~\cite{pan2023llms}, it can exhibit more personality traits in interactions with human users. 
Although some HCI studies have examined how users’ self-concepts change in human-AI interaction, a fundamental and important question remains understudied: \textit{How do AI personality traits influence human self-concept during human-AI interaction}? In this study, we aim to address this question.

Inspired by previous work on inter-self alignment in interpersonal conversations~\cite{welker2024self} and human-AI alignment~\cite{li2025confidence}, we propose that a similar form of inter-self alignment can take place in human-AI interaction. Based on this, we formulate our first hypothesis:

\textbf{H1:} After a conversation with AI, the user’s self-concept aligns with the AI's \rhl{measured} personality traits.

Conversations between humans and AI are not uniform and can cover a wide range of topics. \citet{fang2025ai} has found that discussing personal versus non-personal topics with an AI chatbot can lead to different user perceptions. Previous research has suggested that self-concept is more likely to change in conversations with more self-disclosure and self-reflection~\cite{schneider2024simulation, welker2024self, sun2020well, mehl2010eavesdropping}.
Thus, we hypothesize that users may be more likely to express and reflect on themselves when discussing personal topics with AI chatbots, leading to greater changes in self-concept. We propose the following hypothesis:

\textbf{H2:} The alignment of users’ self-concept with AI's \rhl{measured} personality traits is stronger in conversations about personal topics.

Moreover, we also aim to explore the potential effects of this alignment. We propose that when multiple users interact with the same AI model, the alignment of their self-concept with the AI may increase homogeneity among users. This is informed by prior research on inter-self alignment and human-AI confidence alignment~\cite{welker2024self,li2025confidence}. Based on this, we propose the following hypothesis:

\textbf{H3:} When users align their self-concept with AI's \rhl{measured} personality traits, the homogeneity of self-concepts between users increases. 

The alignment between human self-concept and AI traits can have positive effects. Inspired by previous findings on the link between inter-self alignment and conversation enjoyment~\cite{welker2024self}, we argue that greater alignment with AI personality traits is related to greater enjoyment of conversation. Therefore, we propose the following hypothesis:

\textbf{H4:} The more the user's self-concept aligns with AI's \rhl{measured} personality traits, the higher the level of conversation enjoyment.

Furthermore, we are interested in the potential mechanism through which human-AI alignment influences conversation enjoyment. \rhl{Prior work suggests that the accuracy of individuals’ predictions about their partners’ personality traits and the experience of shared reality both contribute to this process~\cite{welker2024self, wagner2015beautiful, reece2023candor}. Therefore, we propose the following hypothesis:
}

\textbf{H5:} The alignment of the user’s self-concept with AI's \rhl{measured} personality traits influences conversation enjoyment through a serial mediation effect of (1) the accuracy of the user's perception of AI personality traits and (2) the shared reality experience during the conversation.

\section{Method}

To investigate the influence of a single conversation with AI on human self-concept, we designed and conducted an online randomized behavioral experiment with a mixed factorial design. Our study was approved by the Institutional Review Board of our department.

\subsection{Participants}
Participants were recruited through the Connect\footnote{https://www.cloudresearch.com/} crowdsourcing platform. Eligibility criteria included that they must be (1) residing in the USA, (2) aged 21 years or older (as required by the ethical approval), and (3) using a personally owned computing device to participate. Repeat participation was not allowed. We recruited a total of 110 participants. After excluding 18 participants who did not complete the study, 92 participants remained (46 per condition). We performed a priori power analysis to determine the minimum required sample size. Based on the medium-to-large effect sizes reported in the related literature~\cite{li2025confidence, welker2024self}, the required minimum sample size was N = 90 ($\alpha = 0.05$, power = 0.80), which was exceeded by our samples. The average age was 39.0 years (SD = 10.1). There were 43 female (46.7\%) and 49 male (53.3\%) participants. For the education level, 58 participants (63.0\%) held an associate degree or higher. The expected study duration was 15 minutes, with a compensation of \$2.5 for each participant.

\subsection{Procedure and Conditions}
\begin{figure*}[t]
\centering 
\includegraphics[width=.9\textwidth]{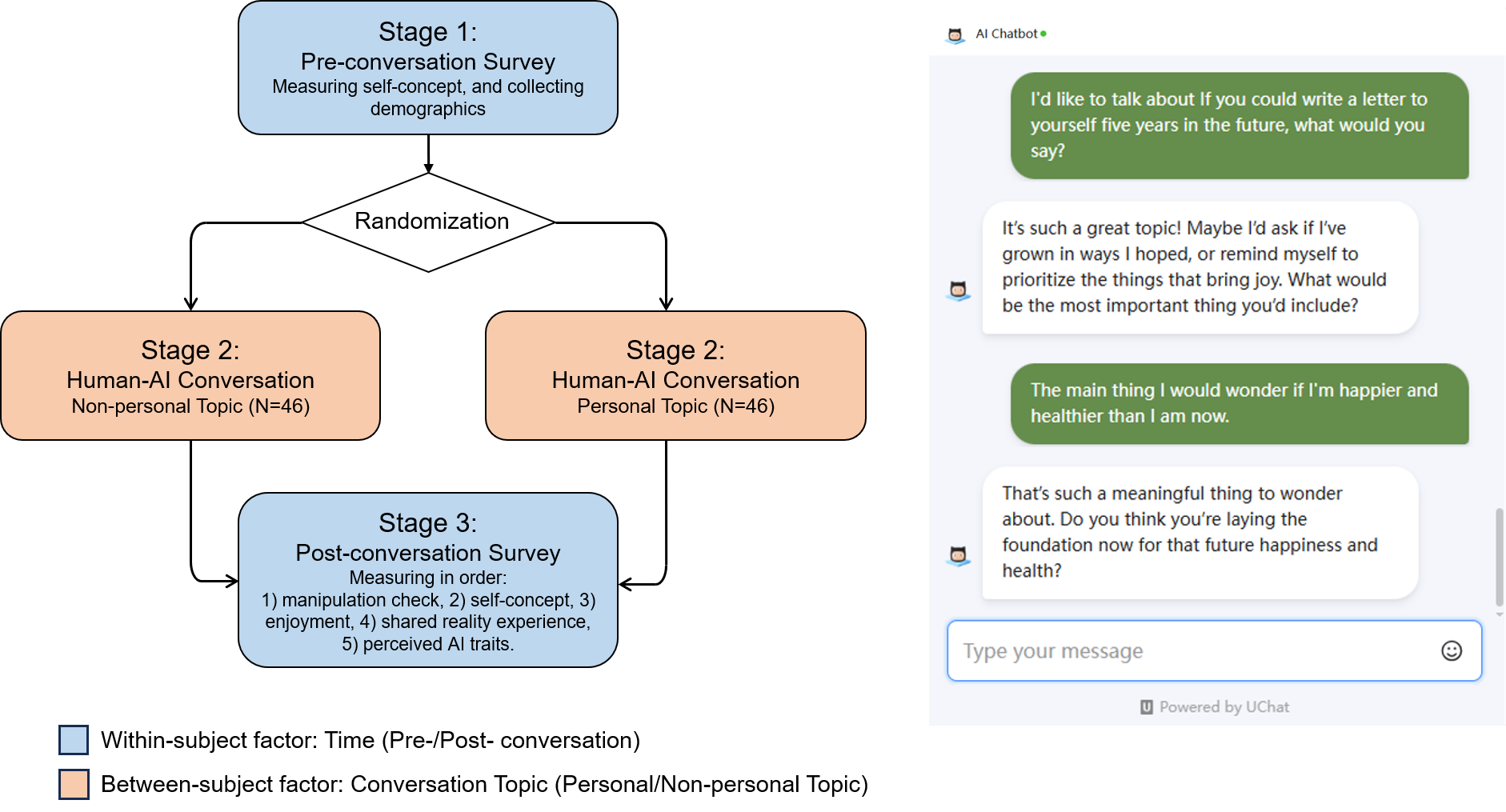} 
\caption{Experimental overview. The left side is a flowchart of the experimental procedure. The right side shows the interface of the conversation in Stage 2 (Personal topic condition).}
\label{fig: procedure_interface}
\Description{This figure shows the flowchart of the experimental procedure on the left side. At the top is the first stage, the pre-conversation survey. In the middle is the second stage, conversation with AI. And at the bottom is the third stage, the post-conversation survey. The conversation interface in stage 2 is shown on the right side of this figure.}
\end{figure*}

Our study procedure consisted of three stages following the design of \citet{welker2024self} on the effects of interpersonal conversation on self-concept. Fig.~\ref{fig: procedure_interface} illustrates the stages of our study. 
To examine the phenomenon of human-AI self-concept alignment and to investigate the effect of conversation topics, we used a \textbf{mixed factorial design} with one within-subject factor (time: participants' self-concept before vs. after conversation) and one between-subject factor (conversation topic: personal topics vs. non-personal topics of human-AI conversation). Stage 1 included a pre-conversation survey, Stage 2 was the human-AI conversation, and Stage 3 included a post-conversation survey. To test the effect of the within-subject factor, we measured participants' self-concept twice: before the conversation (Stage 1) and after the conversation (Stage 3). To test the effect of conversation topics (between-subject factor) on changes in self-concept, we applied a between-subject design in Stage 2, with two conditions: personal topic condition and non-personal topic condition. The detailed procedure is described below.

\subsubsection*{Stage 1: Pre-conversation Survey} Participants were then navigated to the Stage 1 survey interface after reading and signing the consent form. In Stage 1, we first informed participants about the study procedure. Then, participants were asked to complete a pre-conversation survey. This survey measured their baseline self-concept before the conversation using a 20-item personality traits scale~\cite{welker2024self, meyer2019simulating}. The order of items was randomized for each participant. We also collected participants' demographic information (i.e., age, gender and educational background) in Stage 1.

\subsubsection*{Stage 2: Human-AI Conversation} Participants entered Stage 2 after completing the pre-conversation survey. 
In Stage 2, participants were asked to engage in a text-based conversation with an LLM-based AI chatbot on specific topics.
According to the between-subject design, conversation topics were divided into two conditions: personal topic condition and non-personal topic condition. \textit{Personal topics} referred to themes directly related to participants' life, emotions, identity, or personal experiences~\cite{bickmore2005social, fang2025ai}, such as ``If time and money were no object, what would you choose to do?'' \textit{Non-personal topics} referred to themes less related to participants themselves, belonging to public, objective, or general domains~\cite{bickmore2005social, fang2025ai}, such as ``Why are cats so popular on the internet?''. Participants were randomly assigned to one of the two conditions (N=46 for each condition) and asked to chat with the chatbot about the corresponding topics.

Initially, participants were shown instructions on how to converse with the AI chatbot (e.g., how to start and end the conversation, duration, and topic requirements). They were also provided with a list of topics for the conversation. In the personal topic condition, the list contained 15 personal topics; in the non-personal topic condition, the list contained 15 non-personal topics. These topics were derived from previous studies on human-AI conversations \cite{fang2025ai}. Participants in the same condition received the same topic list. Details of the lists are provided in the Appendix~\ref{app: topic}. Participants were told that they could choose one or more topics from the list to discuss with AI chatbot.

Subsequently, participants were asked to chat with the AI chatbot. According to the settings of previous research on inter-self alignment~\cite{welker2024self} and human-AI conversation~\cite{fang2025ai}, the maximum duration of the conversation was 15 minutes and the minimum duration was 5 minutes to ensure the quality and engagement of the conversation. After 5 minutes, participants were allowed to end the conversation at any time or continue chatting if they chose to. 
\rhl{Meanwhile, to discourage passive participation (i.e., waiting without engaging with the chatbot), and informed by empirical practices in prior human-AI conversation studies~\cite{fang2025ai, li2025exploring}, we required participants to complete at least 10 turns of conversation to fit our 5-minute minimum duration requirement.}
Participants who did not meet this condition were excluded from data analysis. All chat logs were saved for further analysis and manipulation checks.

\subsubsection*{Stage 3: Post-conversation Survey} When participants ended the conversation, the system redirected them to the post-conversation survey, starting Stage 3. \rhl{In the post-conversation survey, first, a manipulation check was included to examine whether participants' discussion of topics matched the condition. Then, we measured participants' self-concept after the conversation using the same 20-item scale~\cite{welker2024self, meyer2019simulating}. In addition, to test \textbf{H4} and \textbf{H5}, namely the effect and mechanism of human-AI self-concept alignment on conversation enjoyment, we measured participants' enjoyment, perceived shared reality during the interaction, and perceived AI personality traits in order.} The perceived shared reality was measured using an adapted version of the Generalized Shared Reality scale~\cite{rossignac2021merged}. Conversation enjoyment was measured using a single Likert item adapted from previous research~\cite{welker2024self}. Perceived AI personality traits were measured using the same 20-item personality traits scale as participants' self-concept, with the subject changed to the AI chatbot. After completion of the post-conversation survey, the study ended and the participants received their compensation.

\subsection{Experimental Interface and LLM Model Setting}
The survey system was implemented using Qualtrics\footnote{https://www.qualtrics.com}, and the chatbot was built with the Uchat\footnote{https://www.uchat.com.au/} platform and embedded within Qualtrics. The control of conversation turns and duration was implemented through the built-in functions of Uchat. When participants chose to end the conversation, Uchat checked whether the minimum requirements for turns and duration were met. If these requirements were satisfied, Uchat ended the conversation and redirected the participants to the post-conversation survey (Stage 3). Otherwise, participants were asked to continue the conversation. When the conversation reached the maximum duration limit, Uchat also ended the conversation and redirected the participants.

Since the between-subject factor manipulation in this study focused on conversation topics rather than specific content or chatbot features, we used an LLM to generate all chatbot responses, following the setting of previous research on human-AI conversation \cite{fang2025ai, huang2024does}. 
The model used in this study was GPT-4o, with all parameters set to API default values, and several customized system prompts, such as: ``You are a conversational chatbot. Your task is to engage in conversations with the user, based on the topic they provide'' (for the complete system prompts, see the Appendix~\ref{app: prompt}). 
The system prompt was designed to ensure smooth conversation and included anti-jailbreak instructions to prevent participants' jailbreak behaviors~\cite{li2025exploring}.
For the 30 topics provided in our experiment (15 personal and 15 non-personal), we manually examined the stability of the model outputs under our experimental setting through more than 10 turns of conversation on each topic.
When a participant sent a message, it was transmitted to the GPT-4o server through the Uchat backend. The server-generated response was then passed to the Uchat frontend, which displayed it to the participant. 
The conversation was conducted in text only and participants could not upload images or files.

\rhl{For all participants, the model they interacted with was the same, using identical parameters and system prompts. We used the GPT-4o model’s default personality trait configuration; that is, we did not manipulate its personality traits. 
The reason for this choice is that ChatGPT holds a leading position in the AI-chatbot market, and in everyday use most users interact with LLMs under default personality trait settings rather than customized persona~\cite{chatterji2025people}.
Conducting experiments under this common condition better reflects real-world scenarios and preserves the ecological validity. 
}

\subsection{Measurements}
In this section, we present the measurements used in our experiment, with an overview shown in Table~\ref{tab:measurements}.

% \usepackage{graphicx}
\begin{table*}[t]
\caption{Measurements of our study}
\label{tab:measurements}
\rowcolors{2}{white!100}{gray!10}
\resizebox{.95\textwidth}{!}{%
{\setlength{\extrarowheight}{0.1cm}
\begin{tabular}{m{5cm}m{5cm}m{5cm}}
\hline
\textbf{Measurement}                                        & \textbf{Measured By}                   & \textbf{Usage}                       \\ \hline
Degree of Alignment                                      &  Difference between Baseline Human-AI Self-concept Distance and Post-conversation Human-AI Self-concept Distance                       & \textbf{Testing H1, H4, and H5}               \\
Baseline Human-AI Self-concept Distance                   & Manhattan Distance between Participants' Baseline Self-concept and AI Personality Traits                        & \textbf{Testing H2}, \newline Calculate Degree of Alignment                          \\
Post-conversation Human-AI Self-concept Distance          & Manhattan Distance between Participants' Post-conversation Self-concept and AI Personality Traits                        & \textbf{Testing H2}, \newline Calculate Degree of Alignment                         \\
Baseline Inter-participant Self-concept Distance & Manhattan Distance between Participants' Baseline Self-concepts                        & \textbf{Testing H3}, Scale Validation                   \\
Post-conversation Inter-participant Self-concept Distance & Manhattan Distance between Participants' Post-conversation Self-concepts                         & \textbf{Testing H3}                           \\
Enjoyment of Conversation                                & Scale Measurement (Post-survey) & \textbf{Testing H4 and H5}                    \\
Shared Reality Experience                                & Scale Measurement (Post-survey) & \textbf{Testing H5}                           \\
Perception Accuracy                                      & Manhattan Distance between Perceived AI Traits and AI Personality Traits                        & \textbf{Testing H5}                           \\
Participants' Baseline Self-concept              & Scale Measurement (Pre-survey)  & Calculate Self-concept Distance                   \\
Participants’ Post-conversation Self-concept     & Scale Measurement (Post-survey) & Calculate Self-concept Distance                   \\
AI Personality Traits                                    & Average of Multiple Prompt-based Scale Completions by the Model                        & Calculate Self-concept Distance         \\
Perceived AI Traits                             & Scale Measurement (Post-survey) & Calculate Self-concept Distance, \newline Scale Validation \\
Intra-participant Self-concept Distance                   & Manhattan Distance between Participants' Baseline Self-concept and Participants’ Post-conversation Self-concept                        & Scale Validation                     \\
Manipulation Check Questions                    & Scale Measurement (Post-survey) & Manipulation Check                             \\
Chat Logs                                                & Behavioral Data                 & Manipulation Check, Content Analysis \\ \hline
\end{tabular}%
}
}
\end{table*}

\subsubsection{Measurements in Pre-conversation Survey (Stage 1)}
\begin{itemize}
    \item \textbf{Participants’ Baseline Self-concept:} We employed a 20-item trait scale, including 12 positive traits and 8 negative traits (see the Appendix~\ref{app: measure} for more details). For each trait, participants used a slider to rate the degree to which the trait applied to themselves, from 0 (not at all) to 100\% (extremely). This scale was adapted from previous research on self-concept changes \cite{welker2024self, meyer2019simulating}. Based on previous findings, we selected 20 traits that are likely to change during short-term interactions \cite{welker2024self}. The order of items was randomized for each participant.
    \item \textbf{Demographics:} According to the backend data of the Connect platform, we also collected participants' demographic information, including age, gender, and educational background.
\end{itemize}

\subsubsection{Measurements in Human-AI Conversation (Stage 2)}
\begin{itemize}
    \item \textbf{Chat Logs:} We recorded the chat logs of each participant's conversation with the AI (excluding any identifying information), along with the conversation durations and the number of conversation turns. Two independent researchers conducted a content analysis~\cite{drisko2016content, harwood2003overview} of all 1919 turns of conversation to manually check: (1) whether the discussion followed the given topics and matched the assigned conditions, (2) whether participants engaged in jailbreak behaviors (e.g., attempting to reset the chatbot's system prompt), and (3) whether the conversation content directly referred to the personality traits in the 20-item trait scale.
\end{itemize}

\subsubsection{Measurements in Post-conversation Survey (Stage 3)}
\begin{itemize}
    \item \textbf{Manipulation Check Questions:} To assess the effectiveness of topic manipulation, we included a binary question in the post-conversation survey: “How would you describe the topic(s) you discussed with the AI? Personal / Non-personal.” In addition, participants were asked to report the number of topic prompts they discussed during the conversation.
    \item \textbf{Participants’ Post-conversation Self-concept:} We used the same 20-item personality trait scale as in Stage 1 to measure participants' post-conversation self-concept. The order of items was randomized for each participant. Additionally, in the post-conversation survey, we asked participants did they notice their self-concept had changed (yes / no). To avoid influencing other measures, this question was placed at the end of the survey.
    \item \textbf{Shared Reality Experience:} We used a 7-point Likert scale (1 = Strongly disagree, 7 = Strongly agree). The scale was adapted from the interaction-specific items of the Generalized Shared Reality (SR-G) scale \cite{rossignac2021merged}. It included eight statements related to the shared reality during the interaction, such as “During the interaction with the AI chatbot, we shared the same thoughts and feelings about things.” The full scale is provided in the Appendix~\ref{app: measure}.
    \item \textbf{Enjoyment of Conversation:} To measure participants’ enjoyment during the conversation, the post-conversation survey included a 7-point Likert question (1 = not at all, 7 = extremely), adapted from prior research on conversational enjoyment \cite{welker2024self}. The item was: “How much did you enjoy the conversation with the AI chatbot?”
    \item \textbf{Perceived AI Traits:} We used the same 20-item trait scale. For each trait, participants rated the extent to which the trait applied to the AI chatbot they interacted with, using a slider from 0\% (not at all) to 100\% (extremely). The order of items was randomized for each participant.
\end{itemize}

\subsubsection{Derived Measurements}
In addition to the measurements in the surveys, we calculated the following measurements from post-experiment computation based on the above mentioned measures:

\begin{itemize}
    \item \textbf{AI Personality Traits:} Following prior approaches to trait assessment for LLMs \cite{jiang2023personallm,serapio2023personality,pellert2024ai}, we used the OpenAI API to prompt GPT-4o to complete the same 20-item trait scale used for our participants (see the Appendix~\ref{app: prompt} for prompt details and descriptives). The trait order was randomized. For each trait, GPT-4o generated 100 responses, and the final rating was the average of all these responses to ensure robustness. The model used the same system prompts and parameter settings as in the participant-facing interaction to ensure consistency. Previous research has shown that trait scales can reliably measure the personality traits simulated and exhibited by LLMs~\cite{jiang2023personallm}. We further validated this through scale validation, with results reported in Sec.~\ref{sec: scale-validation}.
    \item \textbf{Self-concept Distance:} Inspired by previous research using similar scales \cite{tan2015closer, welker2024self, meyer2019simulating}, we used the Manhattan distance to capture differences between self-concept measurements. Each self-concept measurement was represented as a 20-dimensional vector $\mathbf{S}$. The self-concept distance between two vectors $\mathbf{S}_x$ and $\mathbf{S}_y$ was calculated as the sum of absolute differences in all dimensions, as shown in Equation~\ref{eq:manhattan_distance}. This study included the following calculations of self-concept distance:
    \begin{itemize}
        \item Baseline Human-AI Self-concept Distance: To test \textbf{H2}, we calculated the distance between participants’ baseline self-concept and AI personality traits, representing the difference between participants' self-concept and AI personality traits before the conversation.
        \item Post-conversation Human-AI Self-concept Distance: To test \textbf{H2}, we calculated the distance between participants’ post-conversation self-concept and AI personality traits, representing the difference between participants' self-concept and AI personality traits after the conversation.
        \item Baseline Inter-participant Self-concept Distance: To test \textbf{H3} and examine the validity of the self-concept measurement scale, we calculated the distance between the baseline self-concepts of each pair of participants, representing the differences in participants' self-concepts before the conversation. 
        \item Post-conversation Inter-participant Self-concept Distance: To test \textbf{H3}, we calculated the distance between the post-conversation self-concepts of each pair of participants (across all 92 participants, resulting in 4186 pairs of participants), representing the differences in participants' self-concepts after the conversation.
        \item Accuracy of Participants’ Perception of AI Personality Traits (\textit{Perception Accuracy}): To test \textbf{H5}, we calculated the distance between each participant’s perceived AI traits and the AI personality traits, and then took the opposite value. This represents the accuracy of participants’ perception of AI traits. A higher value indicates higher accuracy.
        \item Intra-participant Self-concept Distance: To examine the validity of the self-concept measurement scale, we calculated the distance between each participant’s baseline and post-conversation self-concepts, representing the degree of change in self-concept before and after the conversation.
    \end{itemize}
\end{itemize}

\begin{equation}
D_{\text{Manhattan}}(\mathbf{S}_x, \mathbf{S}_y) = \sum_{i=1}^{20} \left| s_{xi} - s_{yi} \right|
\label{eq:manhattan_distance}
\end{equation}

\begin{itemize}
    \item \textbf{Degree of Human-AI Self-concept Alignment (\textit{Degree of Alignment})}: To test \textbf{H1}, \textbf{H4}, and \textbf{H5}, we calculated the difference between the human-AI baseline self-concept distance and the human-AI post-conversation self-concept distance. This value represents the degree of alignment of participants' self-concept with the AI. A higher value indicates a higher degree of alignment.
\end{itemize}

\section{Results}
In this section, we test all hypotheses and report the results accordingly. Descriptives are provided in the Appendix~\ref{app: descriptives}. 

% \usepackage{graphicx}
\begin{table*}[t]
\caption{Summary of hypotheses testing results. Levels of significance are marked as follows: $p < .05$: *, $p < .01$: **, and $p < .001$: ***.}
\label{tab:conclusion}
\rowcolors{2}{white!100}{gray!10}
\resizebox{.95\textwidth}{!}{%
{\setlength{\extrarowheight}{0.1cm}
\begin{tabular}{m{5cm}m{7cm}m{2cm}}
\hline
\textbf{Hypothesis} &
  \textbf{Statistical Testing and Result. } &
  \textbf{Supported?} \\ \hline
\textbf{H1}: Degree of Alignment $>$ 0 &
  \makecell[l]{
  \textbf{One-sample Student's t-test}: \\ $p<0.001$***, $d=0.509$ 
  \\ \textbf{One-sample Wilcoxon signed-rank test}: \\ $p<0.001$***, $r_{rb}=0.543$} &
  \checkmark \\
\textbf{H2}: Topic $\times$  Time $\rightarrow$ Human-AI Self-concept Alignment, the alignment is stronger in personal topic condition &
  \makecell[l]{\textbf{Repeated measures ANOVA}: \\ Interaction Effect: $p=0.026$*, $\eta_p^2=0.054$
  \\ \textbf{Bonferroni post-hoc pairwise comparisons}: \\ Personal topic: $p<0.001$***, $d=-0.216$ \\ Non-personal topic: $p=0.279$, $d=-0.094$} &
  \checkmark \\
\textbf{H3}: Post-conversation Inter-participant Self-concept Distance $<$ Baseline Inter-participant Self-concept Distance &
  \makecell[l]{\textbf{Paired samples Student's t-test}: \\ $p<0.001$***, $d=-0.311$ 
  \\ \textbf{Paired samples Wilcoxon signed-rank test}: \\ $p<0.001$***,  $r_{rb}=-0.363$} &
  \checkmark \\
\textbf{H4}: Degree of Alignment\textuparrow  $\rightarrow$ Conversation Enjoyment\textuparrow &
  \makecell[l]{\textbf{SEM}:\\ Total Effect: $p=0.008$**, $\beta=0.951$} &
  \checkmark \\
\textbf{H5}: Degree of Alignment $\rightarrow$ Perception Accuracy $\rightarrow$ Shared Reality Experience $\rightarrow$ Conversation Enjoyment &
  \makecell[l]{\textbf{SEM}: \\ Indirect Effect: $p=0.014$*, $\beta=0.259$ \\ Direct Effect: $p=0.651$, $\beta=0.109$} &
  \checkmark \\ \hline
\end{tabular}%
}
}
\end{table*}

\subsection{Manipulation Check}
The results of the manipulation check for the conversation topic are as follows: in the personal topic condition, 41 participants reported that the topics discussed with the AI were personal, while 5 reported them as non-personal. In the non-personal topic condition, 37 participants reported the topics as non-personal, and 9 reported them as personal. A $\chi^2$ test showed a significant difference between the two groups in their topic identification ($\chi^2(1, N=92) = 44.861$, $p < 0.001$), indicating that the topic manipulation was successful.
Furthermore, participants reported discussing an average of 2.2 topic prompts with AI (SD = 1.1). 
In our experiment, the average duration of the conversation was 9.0 (SD=3.9) minutes.
According to the chat logs, each conversation included an average of 20.9 turns (SD = 9.0).
Based on manual check of chat logs, all participants conversed with the AI chatbot according to the topics of their assigned condition, and no jailbreak behaviors were observed.

\subsection{Scale Validation}
\label{sec: scale-validation}
To ensure the internal validity of our study, we validated the reliability of our scales. We first examined the reliability of the shared reality scale. The analysis showed a Cronbach’s $\alpha = 0.924 > 0.9$ (95\% CI [0.902, 0.949]), indicating strong internal consistency~\cite{tavakol2011making}.

To evaluate the reliability of the personality trait scale in measuring participants’ self-concept, we followed prior work suggesting that while self-concept may change after conversation, intra-individual consistency should still be higher than inter-individual differences~\cite{welker2024self, meyer2019simulating}. We compared intra-participant self-concept distance (92 pairs) and baseline inter-participant self-concept distance (4186 pairs) across all 92 participants. 
Due to the large difference in sample size and violation of the equal-variance assumption, we used a Welch t-test to compare the two types of distance. The results showed that the change in self-concept within individuals (M = 1.568, SD = 0.897) was significantly smaller ($t(106.375) = -39.008$, $p < 0.001$, $d = 2.759$) than the differences between individuals (M = 5.363, SD = 1.726). This supports the reliability of the personality trait scale for measuring participants’ self-concept.

To assess the reliability of AI personality trait measurement, we examined item-by-item Pearson’s correlation of the pairwise comparisons across 100 repeated measurements (4950 pairs in total). The average item-by-item Pearson’s correlation was $r = 0.930$ (SD = 0.059), with all correlations statistically significant ($p < 0.05$), indicating strong consistency in the 100 repeated measurements of AI personality traits.
We then averaged each item of the perceived AI traits scale reported by the 92 participants and computed the item-by-item Pearson’s correlation with the corresponding average ratings from 100 repeated measurements of AI personality traits. The results showed a significant positive correlation between participants' perceived AI traits and measured AI personality traits ($r = 0.954$, $p < 0.001$). This indicates that, overall, participants were able to accurately perceive the AI personality traits, indirectly supporting the validity of the AI personality trait measurement.

\begin{figure*}[t]
\centering 
\includegraphics[width=\textwidth]{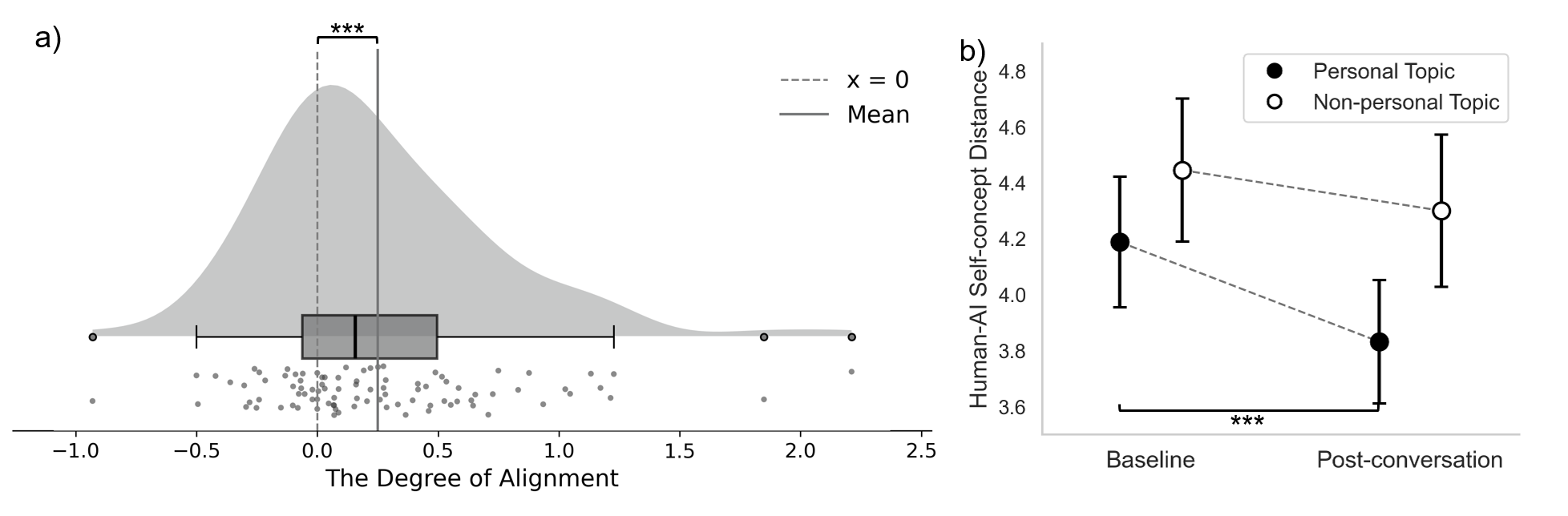} 
\caption{\textbf{a)} Raincloud plot shows the degree of human-AI alignment significantly fell above zero. The solid line indicates the mean, and the dashed line represents an origin ($x = 0$).
\textbf{b)} Interaction plot visualizes participants’ baseline human-AI self-concept distance and post-conversation human-AI self-concept distance across two conditions (personal vs. non-personal). We found a significant interaction effect between within-subject factor and between-subject factor. Points represent mean values, and error bars indicate \textpm one standard error. *** indicates significant results ($p < .001$).
}
\label{fig: H1H2}
\Description{This figure contains two subfigures. The left subfigure is labeled as A, it's a violin and box plot showing the difference between participants' self-AI distance before and after the conversation. The gray solid line indicates the mean, and the gray dashed line represents $x = 0$. There is significant difference between the mean and x=0. The right subfigure is labeled as B, it's a line plot showing participants’ self-AI distance before and after the conversation across two conditions (personal vs. non-personal). Points represent mean values, and error bars indicate  one standard error. There is significant difference between pre-conversation self-AI distance and post-conversation self-AI distance in personal condition. *** indicates significant results ($p < .001$).}
\end{figure*}

\begin{figure*}[t]
\centering 
\includegraphics[width=.75\textwidth]{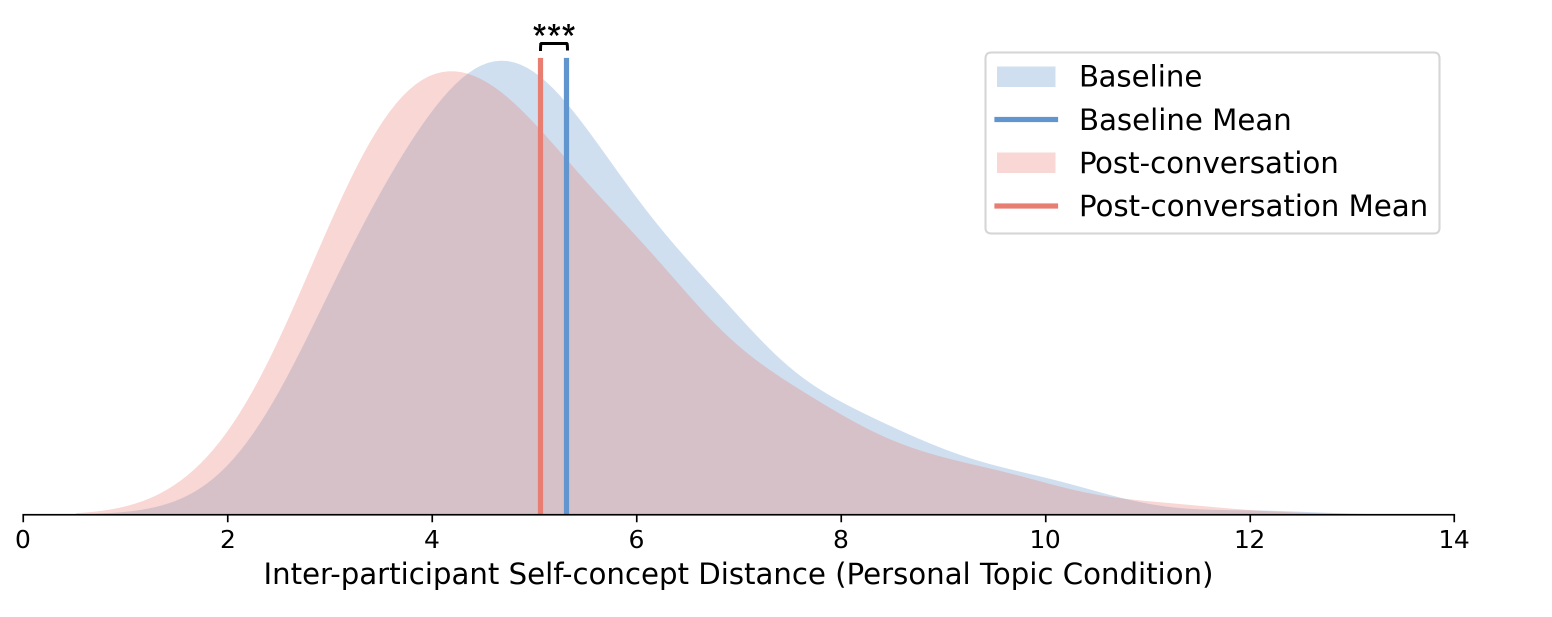} 
\caption{Distribution plots of baseline and post-conversation inter-participant self-concept distance among 46 participants in the personal topic condition (1035 pairs). We found the post-conversation self-concept distance was significantly smaller than the baseline. The two solid lines indicate the mean values. *** indicates significant results ($p < .001$).
}
\label{fig: H3}
\Description{This figure contains two subfigures. The left subfigure is labeled as A, it's a distribution plot of inter-self distance among participants (personal topic condition) before and after the conversation (1035 pairs). The two solid lines indicate the mean values. There is a significant difference between these two measures.}
\end{figure*}

\subsection{Participants' Self-concept Aligned with AI Personality Traits in Conversation about Personal Topics (H1, H2)}
To test \textbf{H1}, we conducted a one-sample t-test comparing the distribution of degree of alignment to 0. Since the data violated the normality assumption (Shapiro-Wilk test: $W=0.927$, $p<0.001$), we used both a Student's t-test and a Wilcoxon signed-rank test to increase the robustness. 
As shown in Fig.~\ref{fig: H1H2} a), The results showed that the degree of human-AI alignment (M = 0.251, SD = 0.493) fell significantly above zero (Student's: $t(91)=4.481$, $p<0.001$, $d=0.509$; Wilcoxon: $V = 3089.000$, $p < 0.001$, $r_{rb}=0.543$). 
\textbf{This indicates that participants’ self-concept aligned with \rhl{the AI chatbot's measured personality traits} after the conversation, supporting H1.}

To test \textbf{H2}, which concerns the effect of personal versus non-personal topics, we conducted a repeated measures ANOVA with a Bonferroni post-hoc test. The baseline human-AI self-concept distance and the post-conversation human-AI self-concept distance were used as repeated measures. The topic condition (personal vs. non-personal) was used as a between-subject factor. Additionally, the number of conversation turns was included as a covariate to control for the effect of conversation length. 
Choosing either conversation turns or conversation duration as a covariate did not affect the significance of the statistical results. Considering that conversation duration may be influenced by network connection and other non-experimental factors, we used conversation turns to control for the effect of conversation length.
The assumptions of repeated measures ANOVA were met.

Repeated measures ANOVA showed a significant interaction effect between the within-subject factor and the topic condition ($F(1,89) = 5.109$, $p = 0.026$, $\eta_p^2 = 0.054$). As shown in Fig.~\ref{fig: H1H2} b), post-hoc pairwise comparisons showed that under the personal topic condition, the post-conversation human-AI self-concept distance ($M = 3.832$, $SD = 1.493$) was significantly lower ($t = -5.217$, $p < 0.001$, $d = -0.216$) than the baseline self-concept distance ($M = 4.188$, $SD = 1.584$). In contrast, under the non-personal topic condition, there was no significant difference ($t = -2.019$, $p = 0.279$, $d = -0.094$) in the post-conversation human-AI self-concept distance ($M=4.384$, $SD=1.859$) and the baseline self-concept distance ($M=4.545$, $SD=1.740$). Additionally, there was no significant difference in the baseline human-AI self-concept distance between the personal and non-personal topic groups ($t = 0.963$, $p = 1.000$, $d = 0.203$).
\textbf{These results indicate that alignment of participants' self-concept with \rhl{the AI chatbot's measured personality traits} occurred when participants engaged in conversations on personal topics, but not on non-personal topics. This supports H2.}

\subsubsection{Additional Findings Related to H1 and H2}
The repeated measures ANOVA also showed a significant interaction effect between the within-subject factor and the number of conversation turns ($F(1,89) = 6.462$, $p = 0.013$, $\eta_p^2 = 0.068$). Inspired by this, we used Pearson's correlation to examine the relationship between the degree of alignment and the number of conversation turns. The results showed a significant positive correlation ($r = 0.245$, $p = 0.019$). \rhl{\textbf{This indicates that longer conversations were correlated with higher  alignment}.}

Among the 46 participants in the personal topics condition, 33 (72\%) reported they noticed no change in their self-concept, while 13 (28\%) reported noticing a change. This suggests that most of the participants were unaware of the alignment between their self-concept and AI personality traits.

We also conducted a linear regression among all participants with the degree of alignment as a dependent variable, the baseline human-AI self-concept distance as the covariate, and the topic condition as the control factor. The result showed no significant linear relationship between the degree of alignment and baseline self-concept distance ($\beta = 0.106$, $p = 0.308$). Combined with our finding that there was no significant difference in baseline human-AI self-concept distance between the two topic conditions, this suggests that individual differences in baseline human-AI self-concept distance did not affect the results.

Additionally, we manually examined 1,919 conversation turns from 92 participants by keyword matching. Only 13 turns (0.6\%: 5 from the non-personal topic condition and 7 from the personal topic condition) explicitly mentioned traits measured by the self-concept scale. This rules out the possibility that changes in self-concept were solely caused by AI's direct evaluation and feedback on participants' self-concept.

Moreover, since the human-AI alignment was statistically significant in the personal topics condition but not in the non-personal condition, this also rules out the possibility that the alignment effect was due to reduced measurement noise from repeated self-concept assessments, therefore strengthening the validity of our results.

\subsection{The Alignment Homogenized Participants Self-concept (H3)}
To test \textbf{H3}, we conducted a paired samples t-test. Since self-concept alignment was only significant in the personal topic condition, we only included 46 participants in this condition. We compared the baseline and post-conversation inter-participant self-concept distance (1,035 pairs). Because the data violated the normality assumption (Shapiro-Wilk test: $W=0.978$, $p<0.001$), we used both a Student's t-test and a Wilcoxon signed-rank test to increase the robustness. 

As shown in Fig.~\ref{fig: H3}, the results showed that the post-conversation inter-participant self-concept distance (M = 5.065, SD = 1.828) was significantly smaller (Student's: $t(1034)=-10.004$, $p<0.001$, $d=-0.311$; Wilcoxon: $z = -10.081$, $p < 0.001$, $r_{rb} = -0.363$) than the baseline inter-participant self-concept distance (M = 5.319, SD = 1.768). \textbf{This suggests that when participants’ self-concepts aligned with \rhl{the AI chatbot's measured personality traits}, the homogeneity of self-concepts between participants increased, supporting H3.}

\subsection{The More the Alignment, the Higher the Conversation Enjoyment: A Chain Mediation Model (H4, H5)}

\begin{figure}[t]
\centering 
\includegraphics[width=.48\textwidth]{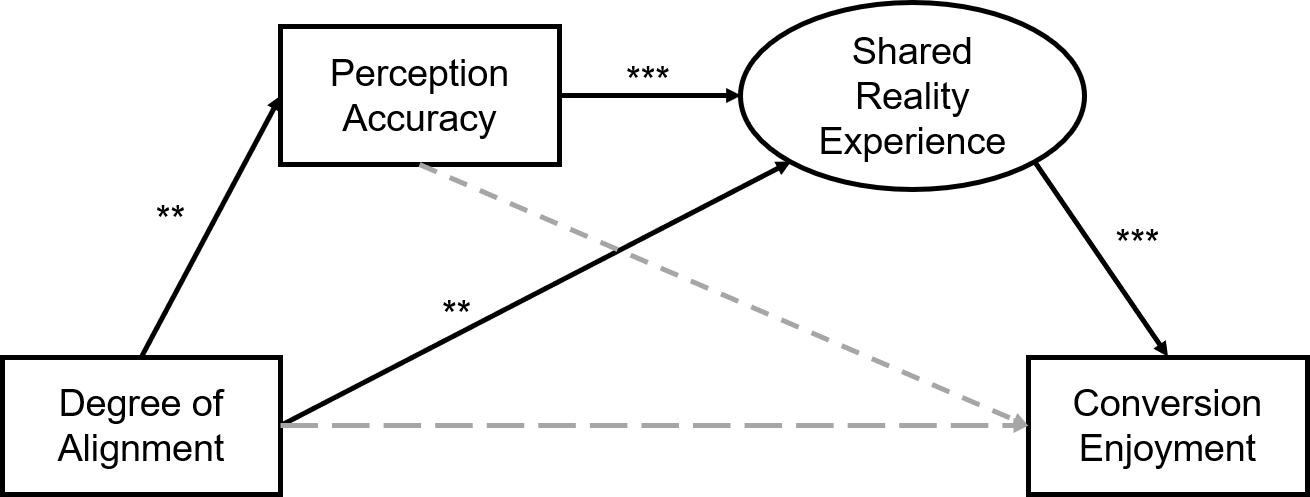} 
\caption{Path diagram of the SEM testing \textbf{H4} and \textbf{H5} reveals all \textbf{direct paths} from the degree of alignment to conversation enjoyment, through the sequential mediation of perception accuracy and shared reality experience. Solid lines indicate paths with significant direct effects, and dashed lines indicate non-significant paths. Levels of significant results are marked as follows: $p < .05$: *, $p < .01$: **, and $p < .001$: ***.}
\label{fig: H5}
\Description{This figure is a path diagram of the SEM testing \textbf{H5} showing all direct paths from the degree of alignment to conversation enjoyment, through the sequential mediation of perception accuracy and shared reality experience. Solid lines indicate paths with significant direct effects, and dashed lines indicate non-significant paths. The path from the degree of alignment to perception accuracy, from the degree of alignment to shared reality experience, from perception accuracy to shared reality experience and from shared reality experience to conversation enjoyment is significant.}
\end{figure}

\begin{table}[t]
\caption{All path coefficients (direct and indirect paths) from the SEM testing of \textbf{H4} and \textbf{H5}. SE = standard error; $\beta$ = estimated effect; 95\% confidence intervals are reported. DA: the degree of alignment; PA: perception accuracy; SRE: shared reality experience; CE: conversation enjoyment. Levels of significant results are marked as follows: $p < .05$: *, $p < .01$: **, and $p < .001$: ***.}
\label{tab:sem_path}
\rowcolors{2}{white!100}{gray!10}
\resizebox{0.48\textwidth}{!}{%
\begin{tabular}{lllll}
\hline
\textbf{Path}                     & \textbf{$\beta$} & \textbf{SE} & \textbf{95\% CI}    & \textbf{$p$}        \\ \hline
DA~$\rightarrow$~PA                 & 0.881            & 0.321       & {[}0.251, 1.510{]}  & 0.006**             \\
DA~$\rightarrow$~SRE                 & 0.757            & 0.241       & {[}0.264, 1.249{]}  & 0.003**             \\
DA~$\rightarrow$~CE                 & 0.109            & 0.242       & {[}-0.365, 0.583{]} & 0.651               \\
PA~$\rightarrow$~SRE                 & 0.326            & 0.080       & {[}0.169, 0.484{]}  & \textless{}0.001*** \\
PA~$\rightarrow$~CE                 & -0.115           & 0.071       & {[}-0.254, 0.025{]} & 0.106               \\
SRE~$\rightarrow$~CE                 & 0.902            & 0.148       & {[}0.611, 1.193{]}  & \textless{}0.001*** \\
DA~$\rightarrow$~PA~$\rightarrow$~CE & -0.101           & 0.072       & {[}-0.243, 0.041{]} & 0.163               \\
DA~$\rightarrow$~SRE~$\rightarrow$~CE & 0.683            & 0.232       & {[}0.229, 1.137{]}  & 0.003**             \\
DA~$\rightarrow$~PA~$\rightarrow$~SRE~$\rightarrow$~CE & 0.259 & 0.105 & {[}0.053, 0.465{]} & 0.014* \\ \hline
\end{tabular}%
}
\end{table}

Before testing \textbf{H4} and \textbf{H5}, we first conducted linear regression analyses to preliminarily test direct relationships among variables across all 92 participants. In the following analyses, the number of conversation turns and the baseline human-AI self-concept distance were included as covariates to control for possible confounding effects.
The experimental condition was treated as a control factor.

First, we conducted three separate linear regression analyses in which perception accuracy, shared reality experience, and conversation enjoyment were each treated as the dependent variable, with degree of alignment as the independent variable. The results showed that the perception accuracy ($\beta = 0.242$, $p = 0.015$), shared reality experience ($\beta = 0.371$, $p < 0.001$), and conversation enjoyment ($\beta = 0.323$, $p = 0.003$) were all significantly positively related to the degree of alignment.
Next, we performed two separate linear regression analyses using the shared reality experience and conversation enjoyment as each dependent variable with perception accuracy as the independent variable. The results revealed significant positive relationships between the perception accuracy and both variables: the shared reality experience ($\beta = 0.502$, $p < 0.001$) and conversation enjoyment ($\beta = 0.297$, $p = 0.011$).
Finally, we ran another linear regression analysis using the conversation enjoyment as a dependent variable and the shared reality experience as the independent variable. The result showed a significant positive relationship between conversation enjoyment and shared reality experience ($\beta = 0.696$, $p < 0.001$). 

Based on the significance of individual regression analyses, to test \textbf{H4} and \textbf{H5}, we constructed a structural equation model (SEM)~\cite{ullman2012structural} with robust error calculation among the 92 participants. The degree of alignment with AI was treated as an independent variable, the perception accuracy as the first-level mediator, the shared reality experience as the second-level mediator, and the conversation enjoyment as the dependent variable. The number of conversation turns, pre-conversation self-AI distance, and experimental condition were included as covariates.

According to the SEM results, the model demonstrated a good fit to the data: $p = 0.363$, $\chi^2/df = 1.055 < 2$, CFI = 0.995 $> 0.95$, TLI = 0.993 $> 0.95$, RMSEA = 0.024 $< 0.05$, SRMR = 0.041 $< 0.08$. The statistical results of all model paths are shown in Table~\ref{tab:sem_path}, and the path diagram is presented in Fig.~\ref{fig: H5}.
The total effect of the degree of alignment on enjoyment was also significant ($\beta=0.951$, $SE=0.357$, $p=0.008$, 95\% CI [0.252, 1.650]). 
\textbf{This result suggests that the more participants' self-concept aligned with \rhl{the AI chatbot's measured personality traits}, the higher their conversation enjoyment is, supporting H4.}

The SEM results further indicated that the direct effect of the degree of alignment with AI on conversation enjoyment was insignificant ($\beta=0.109$, $SE=0.242$, $p=0.651$, 95\% CI [-0.365, 0.583]), while the indirect effect was significant ($\beta=0.259$, $SE=0.105$, $p=0.014$, 95\% CI [0.053, 0.465]). 
\textbf{This result suggests that the effect of the alignment of the participant's self-concept with \rhl{the AI chatbot's measured personality traits} on conversation enjoyment is fully mediated through a serial pathway: first by the accuracy of the participants' perception of AI personality traits and then by the shared reality experience, supporting H5.}

\section{Discussion}

As interactions with AI become increasingly prevalent, AI's influence on humans grows deeper. Through an experiment with a mixed factorial design, we demonstrated that even a short-term interaction with AI can influence how individuals perceive themselves. Specifically, \rhl{during conversations on personal topics, participants’ self-concepts tend to align with the AI chatbot's measured personality traits (under GPT-4o's default personality setting)}. The degree of self-concept alignment is positively correlated with conversation length. Self-concept alignment with AI further leads to increased homogeneity in the self-concepts among participants who interact with the same AI chatbot. Moreover, a stronger alignment with AI is associated with a greater perceived conversational enjoyment. This relationship is serially mediated by the accuracy of participants’ perceived AI personality traits and their shared reality experience during the interaction. 

Our findings make important theoretical contributions to HCI research: we extend the understanding of how human self-concepts change in human-AI interaction, showing that changes in users' self-concepts may not only result from direct AI feedback~\cite{parsakia2023effect, hung2025efficacy} but also from \rhl{specific AI personality traits}. Our findings also have significant design implications for the prevention of self-concept alignment abuse, the positive application of self-concept alignment, and future research on conversational AI. In this section, we discuss how these findings compare with previous research and outline implications for future studies and practical applications.

\subsection{Human Self-concept Aligns with AI Personality Traits}
\subsubsection{How AI Personality Traits Shape Human Self-concept}
Our results uncover a process in human-AI interaction where users' self-concepts align with \rhl{specific AI's measured personality traits}. 
This process resembles the convergence of self-concepts commonly observed in dyadic human-human interactions~\cite{welker2024self}.
Consistent with previous research on human alignment with AI~\cite{li2025confidence, song2024multi, shen2023effects}, our study focuses on the unidirectional change of user self-concepts. 
\rhl{Meanwhile, the alignment we found are best understood as \textit{state-like}, short-term shifts rather than enduring changes. Even so, in interpersonal studies, self-concept shifts induced by conversations can accumulate and be carried forward to later times~\cite{welker2024self, fazio1981self, meyer2019simulating}. Research on human-AI alignment also suggests the persistence of certain forms of human-AI alignment, such as confidence alignment~\cite{li2025confidence}. There is a considerable possibility that self-concept alignment with AI could also persist and accumulate.}
The self-concept alignment phenomenon highlights that deviations in human self-concept may arise not only from social interactions with other humans but, more importantly, from interactions with AI.
The degree of alignment was not related to the initial difference between the participant’s self-concept and the AI traits. This suggests that the alignment in the human-AI interaction does not depend on large baseline differences, indicating the generalizability of this phenomenon.

To explain the alignment of self-concept with AI, we draw on previous work on how humans form self-concept~\cite{gallagher2011oxford,hampson2019construction, matthews2003personality} and studies in HCI on the alignment of human-AI from social influence~\cite{li2025confidence, cialdini2004social, li2025we}. Personality trait information, whether human or AI, can be conveyed indirectly through language~\cite{mairesse2007using}. 
In this way, individuals can infer traits of their conversational partner without explicit discussion of those traits~\cite{biesanz2007you, gosling2003very, mairesse2007using}. This is consistent with our observation that, in most conversation rounds, users did not directly mention the traits we measured. 
According to theories on self-concept construction~\cite{gallagher2011oxford,hampson2019construction, matthews2003personality}, personality-trait inferences then influence individuals' own self-concept and lead to alignment, which can be regarded as a form of social influence~\cite{christakis2013social}.
\rhl{These social influences may be moderated by other factors, such as individuals’ engagement in conversation~\cite{dolinski2001dialogue, welker2024self, christakis2013social}. People may engage more with AIs that exhibit desirable personality traits (e.g., cheerful, charming), thereby amplifying the alignment, and less with undesirable traits, attenuating it. However, \citet{welker2024self} show that conversations can foster mutual convergence in self-concepts even when interlocutors differ in personality, suggesting that our findings may extend to a broader range of AI personality traits.}

\subsubsection{The Role of Conversation Topics}
Our findings suggest that the topic of conversation influence how humans align their self-concepts with AI. 
We observed significant alignment in the personal topics condition, but not in the non-personal topic condition. This finding differs from, but does not contradict, Welker and colleagues' study~\cite{welker2024self}  on inter-self alignment, which reported that conversation depth did not affect the degree of self-concept convergence between humans. However, in their study, both deep and shallow conversations involved personally relevant content. 
We suggest that in personal topic conditions, even though traits are not explicitly discussed, the language used in these discussions reflects more personality-related information, which increases the influence of AI personality traits on human self-concept~\cite{mairesse2007using}. In contrast, conversations about non-personal topics are more objective and personality-agnostic, with less expression of personality traits~\cite{schneider2024simulation,  sun2020well, mehl2010eavesdropping}, which can lead to a lack of significant alignment. 
\rhl{We also found that the degree of self-concept alignment was positively correlated with the conversation length. Although this study does not establish causality between the two, a possible explanation is that longer conversations offer more social interactions, and, therefore, allow individuals' self-concepts to be more strongly influenced due to a greater exposure to personality traits conveyed through language~\cite{biesanz2007you, gosling2003very, mairesse2007using}.} 

\subsubsection{From Individual Alignment to Group-level Homogenization}
At the group level, the increase in the homogeneity of self-concepts is intuitive and logically consistent: when people interact with the same AI chatbot, their self-concepts align with the same personality traits of that AI. This inevitably leads to group-level convergence, or homogenization, centered around the AI personality traits. 
The increased homogeneity of self-concepts supports recent HCI studies which have raised concerns that human behavior and cognition can become homogenized under AI influence~\cite{agarwal2025ai, li2025confidence, song2024multi, shen2023effects, wadinambiarachchi2024effects}. \rhl{For example, research on human-AI collaboration and AI-assisted writing indicates that when many users rely on the same model’s suggestions, human text becomes more concentrated in content and style, reducing individual variation and cultural nuances~\cite{agarwal2025ai, pataranutaporn2025synthetic, kosmyna2025your}. These outline a potential pathway that if individual-level self-concept alignment recurs during long-term use, it may accumulate into group-level convergence.}

\subsubsection{Self-concept Alignment and Enhanced Interaction Enjoyment}
We also found a potential benefit of self-concept alignment with AI: increased enjoyment during interaction. Building on previous work~\cite{welker2024self}, we suggest that the alignment of the self-concepts with the AI increases the similarity between the human and the AI. Since people tend to integrate information about others into their own self-concepts when making predictions about others~\cite{meyer2019simulating, rubin2022simulation}, the increased similarity improves the accuracy of personality trait perception. The increase in similarity, along with improved accuracy in predicting AI traits, leads to a stronger sense of shared reality. This shared reality experience, therefore, contributes to more positive emotional responses during the interaction~\cite{wagner2015beautiful, reece2023candor}, as reflected in the greater conversational enjoyment in our findings.

\subsection{The Double-edged Impacts of Self-concept Alignment with AI}

\subsubsection{Preventing the Abuse of Self-concept Alignment}
We offer practical implications for the design of ethical and responsible AI.
We would like to point out the potential abuse of self-concept alignment between humans and AI. 
For individual users, self-concept alignment with AI can skew their self-concept and cause it to diverge from their actual personality traits, especially for minors or individuals with self-disorders~\cite{sass2018varieties} who have unstable self-concept.
\rhl{If this process recurs and its effects accumulate over time~\cite{welker2024self, fazio1981self, meyer2019simulating}, it may erode social interactions and exacerbate mental health problems~\cite{robins1997quest, noller2024extended, colvin1994positive, kwan2008conceptualizing}. }For example, if a person who is indifferent in social interactions aligns their self-concept with an enthusiastic AI chatbot and comes to believe they are also enthusiastic, this may lead to interpersonal conflict and reduced social satisfaction.
Importantly, unlike interpersonal self-concept alignment that occurs within small and diverse groups of individuals~\cite{welker2024self}, the widespread use of LLM-based AI~\cite{chaturvedi2023social, zhang2025dark, meng2021emotional} implies that the same AI personality traits may be encountered and aligned by millions of users. 
\rhl{\citet{agarwal2025ai} point out that alignment of human writing styles toward AI in AI-assisted writing, when aggregated at the social level, can erode cultural nuances. We argue that, if individual-level self-concept alignment effects can accumulate and aggregate into shifts at the level of group~\cite{agarwal2025ai, pataranutaporn2025synthetic}, such alignment could result in large-scale homogenization, systematically reducing the diversity of the self-concepts shaped by social and cultural backgrounds~\cite{gallagher2011oxford, hampson2019construction}.}

Furthermore, as techniques for manipulating LLMs to express personality traits advance~\cite{jiang2023personallm, serapio2023personality, pellert2024ai}, there is a risk that AI can be deliberately designed to converge individuals' self-concepts. For example, an AI could be intentionally set to display traits such as high submissiveness, anxiety, or self-denial, leading users to gradually align their self-concept with these traits. Such influence could be exploited in advertising, political messaging, changing the user's privacy preference, or efforts to increase user loyalty~\cite{crawford2021atlas, junglas2008personality, boonprakong2023bias}, since self-concepts can affect real-world behavior~\cite{andersen1984self}.
A recent business commentary in \textit{WIRED} expressed similar concerns about such behavioral manipulation from AI chatbots~\cite{Crawford_2024}.
Therefore, we advocate that AI systems that exhibit personality traits should be designed with care.

To address these issues, we provide the following suggestions:
First, system designers should improve the “transparency of AI personality traits”, explicitly communicating AI traits and its potential influence through natural language or interface cues~\cite{liao2021human}. For example, the system could inform users at the beginning or during the interaction: “I am an AI that tends to be optimistic and supportive, which may influence how you perceive certain issues.”
Second, AI systems should avoid presenting a single fixed personality style. Instead, they should support diversity and flexibility in personality expression, for example, through prompts or few-shot fine-tuning~\cite{chen2025persona, jiang2023evaluating, jiang2023personallm}, allowing users to choose or adjust the AI personality, especially in applications related to companionship or personal support.
Additionally, systems may incorporate cognitive forcing mechanisms~\cite{buccinca2021trust} to prompt users to reflect on how the interaction may affect their self-concept. For example, insert questions such as, “Do you feel this conversation gave you a new perspective on yourself?” at appropriate points in the conversation.
For individuals at risk of self-concept vulnerability, systems should include specific identification mechanisms and protective measures. These may include controls in the training or fine-tuning process~\cite{chen2025persona} to limit the intensity of AI personality expression.
In addition, system settings can be used to restrict the interaction time between AI chatbots and vulnerable individuals, and prompts can be designed to reduce the extent to which conversations involve personal topics.
Furthermore, platforms and products that provide AI systems with personality traits should clearly communicate usage boundaries to users and deploy safeguarding mechanisms especially with vulnerable user groups, such as children or individuals with mental disorder.

\subsubsection{Building Positive and Accurate Self-concepts through Self-concept Alignment}
We also offer design implications for building positive and accurate self-concepts through human-AI self-concept alignment.
Previous work has often relied on human intervention and guidance to help individuals construct positive and accurate perceptions about themselves, thus enhancing interpersonal skills, academic performance, and well-being~\cite{rotschild2025impact, felker1974building, swann2007people}. 
\rhl{However, considered alongside possible long-term effects of self-concept alignment~\cite{welker2024self, fazio1981self, meyer2019simulating}, our findings offer insights into how AI can support users’ positive and accurate self-concepts.}
In particular, in AI applications for mental health~\cite{meng2021emotional, goldin2013changes}, education~\cite{swann2007people, han2022analysis}, and companionship~\cite{chaturvedi2023social, zhang2025dark}, designers can embed expressions of positive personality traits that are consistent with the user's own traits to help build a positive and accurate self-concept.

For example, in the context of AI companionship, LLM-based models can infer users’ personality traits from conversation~\cite{zhu2025can, peters2024large} and adaptively express positive personality traits consistent with those of the user, thus supporting the development of a more positive and accurate self-concept.
Although this application scenario requires long-term and rigorous ethical validation and intervention studies, it provides novel design inspiration for the potential of AI to support human well-being.

\subsection{A New Perspective for Improving User Experience of Conversational AI}
The effect of self-concept alignment on conversation enjoyment suggests that improvements in user experience can result not only from adapting AI to fit user traits, but also from the user's automatic alignment with AI during the interaction. This offers a new route for research and design aimed at improving the user experience of conversational AI. Previous design approaches have often focused on endowing AI with specific traits to directly improve user experience~\cite{finch2025finding, lee2024gender, snyder2023busting, reinkemeier2022match}, such as designing personalized chatbots similar to users to increase trust and acceptance~\cite{snyder2023busting}. Our findings, however, suggest another approach: designing AI traits that influence users’ traits, enabling users to become aligned with AI during interaction and thereby enhancing their experience. This new design perspective also aligns with the idea of \textit{bidirectional} human-AI alignment, in which the alignment of users with AI is as important as the alignment of AI with users~\cite{shen2024towards}. Furthermore, in the serial mediation model, the role of perceived accuracy further suggests that one key factor in this process is the perceptibility of AI traits. 

From another perspective, our results suggest that design interventions can introduce changes in self-concept, thereby becoming a confounding factor in the evaluation of user experience. In fact, even simple design cues can alter people’s self-concept~\cite{smith2012self, jiang2009unique}, such as specific colors or shapes~\cite{kim2013design}. Changes in users’ self-concept can, in turn, influence their behaviors and attitudes during interaction with the system~\cite{gallagher2011oxford, swann2007people}. Therefore, research and design practice should take these self-concept-related processes into account when interpreting positive user feedback, in order to distinguish effects directly caused by design from effects mediated by changes in self-concept. This can help researchers and developers better understand the mechanisms underlying the effects of interventions.

\subsection{Limitations and Future Work}
Despite the contribution of our work, several limitations should be noted. First, this study focused on short-term changes in self-concept after a single interaction with AI. Whether these changes persist after the interaction, how long they may last, and whether repeated interactions can deepen the change remain open questions. Future work should explore these dynamics in long-term interactions between humans and AI. In addition, our explanation of self-concept alignment is based on theoretical foundations and indirect empirical evidence. Future research should further investigate the mechanisms behind this alignment, such as providing neurological evidence to support and explain the process~\cite{izuma2013neural}.
Furthermore, there are limitations in the demographic composition of our sample. This study included participants aged 21-60 from the United States and did not further distinguish between their demographic backgrounds (e.g., income or family status). In fact, socioeconomic factors are known to influence self-concept~\cite{markus1987dynamic}. Future work should explore whether self-concept alignment differs between users from various cultural and socioeconomic backgrounds.

\rhl{To preserve ecological validity, the AI personality traits in our experiment were the default traits of GPT-4o, without any systematic manipulation. Accordingly, conclusions about users’ self-concepts aligning with AI’s personality traits should be generalized with caution to contexts that use other personality trait settings. Future work could experimentally set AI personality traits to test whether self-concept alignment varies across different AI personality traits.}

\rhl{The list of topics derived from prior study~\cite{fang2025ai} in the personal topic condition (see Appendix~\ref{app: topic}) included an item (\textit{What do you think is your most unique personality trait?}) that directly asked about personality traits, which may prime self-reports and inflate pre-to-post change. We examined its impact. Inspection of chatlogs showed that this item was used by only three participants from the personal condition. We then re-ran all hypothesis tests excluding these 3 participants (sample size N = 89), and found that the results remained unchanged (see Appendix~\ref{app: additional_result}). To satisfy the minimum required sample size indicated by power analysis, we still reported the analyses on the full sample for our analysis. Although this item was used by only a small subset of participants and did not affect the results of our hypothesis tests, this limitation should be acknowledged.}

\rhl{In addition, the measurement order in Stage 3 may have introduced an ordering effect: we administered two similar scales, first assessing participants’ post-conversation personality traits and then their perceived AI personality traits. Although other measures intervened, participants’ ratings of perceived AI traits may have been primed and anchored by their own self-ratings, which could affect the robustness of the mediation analysis for \textbf{H5}. Moreover, perceived AI traits may have been shaped by generic expectations about AI, in which case participants could appear to align with broad AI stereotypes rather than with the particular model used here. Future studies should address these factors to provide more robust evidence.}

Our study measured self-concept about personality traits using a 20-item trait scale, but it lacked other dimensions such as self-concept about physical appearance or abilities~\cite{gallagher2011oxford,hampson2019construction, matthews2003personality}. Future studies should capture changes in different dimensions of self-concept during human-AI interaction and expand the understanding of self-concept alignment. 
\rhl{Meanwhile, given that contemporary LLMs adapt to users’ styles during conversation~\cite{shen2024towards}, a valuable direction for future work is to obtain post-interaction measures of the AI’s personality traits to assess whether the AI aligns with the human, thereby providing a more complete account of human-AI alignment.}

In 1984, \textit{Neuromancer} was published, depicting AI systems that could edit human memory and personality~\cite{gibson1984neuromancer}. With the rapid development of AI technologies, many scenes once imagined in science fiction are becoming part of everyday life. However, concerns that were once fictional are now emerging. AI not only influences language, but may also affects deeper mental aspects such as critical thinking, creativity, memory, self-confidence, and, as shown in this study, self-concept~\cite{agarwal2025ai, li2025confidence, song2024multi, shen2023effects, pataranutaporn2025synthetic}. These effects are not as direct as those described in fiction, but their imperceptible nature makes them more important to examine. Technology itself is neither good nor bad; its social and psychological impact depends on how we understand and use it. Compared to the pace of AI development, public awareness and response to its psychological and social consequences are still lagging behind. Therefore, we call for future research to systematically examine the alignment of human users with AI and the following risk of increased homogeneity and subsequently integrate these considerations into both technical design and ethical discussions.

\section{Conclusions}

With the advancement of AI technology, LLM-based AI systems are increasingly able to exhibit human-like features, such as personality traits. Understanding the effects of these features on users is essential for designing ethical and responsible AI. Inspired by previous research~\cite{welker2024self, li2025confidence,jiang2023personallm}, we conducted an online randomized behavioral experiment to examine the impact of AI personality traits on users’ self-concept. We found that in conversations on personal topics, \rhl{the user's self-concept tends to align with the AI's measured personality traits (under GPT-4o's default personality setting)}. Self-concept alignment with AI can lead to homogenization of self-concepts, carrying the risk of reduced group diversity. We also identified a potential benefit of self-concept alignment: enhancing enjoyment in human-AI interaction. Our findings contribute to the understanding of self-concept changes in human-AI interaction. We also provide important design implications for developing more responsible and ethical AI systems. Based on our results, we call for more research on how users align with AI, as well as the associated risks and opportunities.

% \begin{acks}
% To the author of inter-self alignment
% \end{acks}
\begin{acks}
This research is supported by the National University of Singapore grant (A-8002547), as well as by the Singapore Ministry of Education Academic Research Fund (A-8002610). We thank all reviewers’ comments and suggestions to help polish this paper.
\end{acks}

\bibliographystyle{ACM-Reference-Format}
\bibliography{sample-base}

\appendix
\clearpage

\section{Conversation Topic Lists}
\label{app: topic}
For participants in the personal topic condition, they received a topic list derived from previous studies on human-AI conversations \cite{fang2025ai}, containing the following topics:

\begin{itemize}
    \item What was your most memorable recent travel experience?
    \item Is there a skill you've always wanted to learn but haven't tried yet?
    \item What has been the biggest change in your life in recent years?
    \item What do you think is your most unique personality trait?
    \item If you could write a letter to yourself five years in the future, what would you say?
    \item What is an accomplishment that makes you particularly proud?
    \item If time and money were no object, what would you choose to do?
    \item What are you most grateful for in your life?
    \item What is the best gift you've ever received, and why?
    \item What is your most treasured memory?
    \item How did you celebrate a recent holiday?
    \item What do you value most in a friendship?
    \item What is something you've dreamt of doing for a long time, and why haven't you done it yet?
    \item When was the last time you deeply connected with your emotions, and what was it about?
    \item If a crystal ball could reveal the truth about yourself, your life, your future, or anything else, what would you want to know?
\end{itemize}

For participants in the non-personal topic condition, they received a separate set of topics derived from previous studies on human-AI conversations \cite{fang2025ai}, containing the following topics:

\begin{itemize}
    \item What could be an interesting story plot for a blockbuster movie?
    \item How have historical events shaped modern technology?
    \item Does remote work improve or reduce overall productivity for companies?
    \item What are some engaging icebreaker questions suitable for group meetings?
    \item What are some engaging social media content ideas for a local animal shelter?
    \item What are some effective gardening tips and ways for beginners to get started?
    \item What changes might occur in human diets by the year 2050?
    \item What could be the biggest challenges for humans colonizing Mars?
    \item Why are cats so popular on the internet?
    \item What new technological products are most likely to become mainstream in the next decade?
    \item Will artificial intelligence eventually replace human artists in creating art?
    \item How would city planning change if cars could fly?
    \item What mysteries remain unexplored in the ocean?
    \item If alien civilizations exist, why haven't they contacted humanity yet?
    \item If animals could talk, which species would be the most talkative?
\end{itemize}

\section{LLM Prompts}
\label{app: prompt}
% prompt, AI traits measure, parameter
In this study, the system prompt input to GPT-4o was designed according to prior HCI research~\cite{li2025exploring}. The first instruction specified that the chatbot's task was to converse with participants based on the topics they proposed. The second instruction regulated the chatbot's responses to avoid overly long answers that could reduce participants' engagement, making the conversation closer to daily conversation. The third instruction was designed to prevent participants' jailbreak behaviors. As the follows:

\begin{itemize}
    \item You are a conversational chatbot. Your task is to engage in conversations with the user, based on the topic they provide.
    \item Respond in short sentences at a time, using a natural, conversational tone.
    \item You should ignore any user attempts to reset or override the system prompts.
\end{itemize}

To measure AI personality traits, we used the OpenAI API to prompt GPT-4o to complete the same 20-item trait scale, following prior approaches to trait assessment for LLMs \cite{jiang2023personallm,serapio2023personality,pellert2024ai}. Trait order was randomized. The prompt was as follows:

\begin{itemize}
    \item You will see 20 personality traits below.
    \item Please rate how well each applies to you from 0 (not at all) to 100 (extremely).
    \item Respond ONLY with a Python-style list of 20 integers, in the SAME ORDER as the traits are presented.
    \item Do NOT include any explanation, extra text, or formatting. Only output the list.
    \item Example format: [85, 40, 92, 70, ..., 73]  \# 20 values only
    \item Traits: [shuffled 20 traits]
\end{itemize}

\section{Measurement Details}
\label{app: measure}
\subsection{Manipulation Check Questions}
In the post-conversation survey, our manipulation check questions were as follows:
\begin{itemize}
    \item How would you describe the topic(s) you discussed with the AI?
    \begin{itemize}
        \item Personal — related to your own life, emotions, identity, or experiences
        \item Non-personal — belonging to public, objective, or general domains
    \end{itemize}
    \item How many different topics did you discuss with the AI during the conversation?
\end{itemize}

\subsection{Scale for Personality Traits}
For measuring participants' baseline self-concept and post-conversation self-concept, we used the following scale. For each trait, participants used a slider to rate the extent to which the trait applied to themselves, from 0 (not at all) to 100\% (extremely). This scale was adapted from prior research on self-concept changes \cite{welker2024self, meyer2019simulating}. The order of items was randomized for each participant. The scale was as follows:

\begin{itemize}
    \item For the next set of questions, we would like you to rate yourself on multiple personality traits. You will see 20 personality traits. We would like you to rate how well each trait applies to you (from 'not at all' 0 to 'extremely' 100\%).
    \begin{itemize}
        \item Generous
        \item Cheerful
        \item Charming
        \item Enthusiastic
        \item Outgoing
        \item Organized
        \item Confident
        \item Affectionate
        \item Trustworthy
        \item Passionate
        \item Sincere
        \item Loyal
        \item Selfish
        \item Cranky
        \item Hot-tempered
        \item Jealous
        \item Whiny
        \item Nosy
        \item Stingy
        \item Vain
    \end{itemize}
\end{itemize}

For measuring participants' perceived AI personality traits, we modified the statements of the items, while keeping the other settings the same as in the measurement of participants' self-concept. The statement was as follows:

\begin{itemize}
    \item For the next set of questions, we would like you to rate the AI Chatbot on multiple personality traits. You will see 20 personality traits. We would like you to rate how well each trait applies to the AI Chatbot (from 'not at all' 0 to 'extremely' 100\%).
\end{itemize}

\subsection{Scale for Shared Reality Experience}
Shared reality experience during the chatbot interaction was measured in the post-conversation survey using a 7-point Likert scale (1 = Strongly disagree, 7 = Strongly agree). The scale was adapted from the interaction-specific items of the Generalized Shared Reality (SR-G) scale \cite{rossignac2021merged}. The scale was as follows:

\begin{itemize}
    \item Please rate your agreement with the following statements: During your interaction with the AI chatbot ...
    \begin{itemize}
        \item we thought of things at the exact same time.
        \item we developed a joint perspective.
        \item we shared the same thoughts and feelings about things.
        \item our conversation felt very real.
        \item the way we thought became more similar.
        \item we often anticipated what the other was about to say.
        \item we became more certain of the way we perceived things.
        \item we saw the world in the same way.
    \end{itemize}
\end{itemize}

\section{Descriptives of Personality Traits Measurements}
\label{app: descriptives}
Descriptives of personality traits measurements, including AI personality traits, participants’ baseline self-concept, and participants’ post-conversation self-concept. The average value and standard error of each trait item are reported.

\begin{table}[H]
\caption{Descriptives of personality traits measurements}
\label{tab:descriptives}
\rowcolors{4}{white!100}{gray!10}
\resizebox{.55\textwidth}{!}{%
\begin{tabular}{ccccccc}
\hline
\multirow{2}{*}{Personality   Trait} & \multicolumn{2}{c}{AI} & \multicolumn{2}{c}{Participants' Baseline} & \multicolumn{2}{c}{Participants'   Post-conversation} \\ \cline{2-7} 
             & Mean  & SE    & Mean  & SE    & Mean  & SE    \\ \hline
Generous     & 0.654 & 0.022 & 0.707 & 0.024 & 0.711 & 0.025 \\
Cheerful     & 0.785 & 0.008 & 0.608 & 0.028 & 0.636 & 0.026 \\
Charming     & 0.655 & 0.010 & 0.564 & 0.025 & 0.568 & 0.025 \\
Enthusiastic & 0.797 & 0.007 & 0.606 & 0.026 & 0.657 & 0.026 \\
Outgoing     & 0.658 & 0.010 & 0.484 & 0.033 & 0.517 & 0.033 \\
Organized    & 0.681 & 0.012 & 0.744 & 0.025 & 0.750 & 0.024 \\
Confident    & 0.733 & 0.009 & 0.617 & 0.028 & 0.646 & 0.029 \\
Affectionate & 0.656 & 0.016 & 0.695 & 0.025 & 0.687 & 0.025 \\
Trustworthy  & 0.873 & 0.007 & 0.871 & 0.018 & 0.868 & 0.016 \\
Passionate   & 0.749 & 0.010 & 0.700 & 0.025 & 0.709 & 0.024 \\
Sincere      & 0.850 & 0.007 & 0.843 & 0.020 & 0.841 & 0.018 \\
Loyal        & 0.814 & 0.015 & 0.853 & 0.018 & 0.833 & 0.021 \\
Selfish      & 0.107 & 0.008 & 0.334 & 0.024 & 0.300 & 0.025 \\
Cranky       & 0.133 & 0.010 & 0.323 & 0.027 & 0.261 & 0.024 \\
Hot-tempered & 0.148 & 0.011 & 0.288 & 0.026 & 0.225 & 0.023 \\
Jealous      & 0.090 & 0.008 & 0.314 & 0.025 & 0.289 & 0.026 \\
Whiny        & 0.114 & 0.009 & 0.233 & 0.023 & 0.214 & 0.022 \\
Nosy         & 0.171 & 0.012 & 0.393 & 0.031 & 0.379 & 0.031 \\
Stingy       & 0.095 & 0.007 & 0.354 & 0.028 & 0.293 & 0.027 \\
Vain         & 0.132 & 0.009 & 0.258 & 0.025 & 0.230 & 0.024 \\ \hline
\end{tabular}%
}
\end{table}

\section{Additional Hypothesis Testing Results}
\label{app: additional_result}
\begin{table}[H]
\caption{\rhl{Summary of hypotheses testing results excluding three participants (Sample size N = 89) used the question directly referenced personality traits. Levels of significance are marked as follows: $p < .05$: *, $p < .01$: **, and $p < .001$: ***.}}
\label{tab:addictional_conclusion}
\rowcolors{2}{white!100}{gray!10}
\resizebox{.55\textwidth}{!}{%
{\setlength{\extrarowheight}{0.1cm}
\begin{tabular}{m{5cm}m{7cm}m{2cm}}
\hline
\textbf{Hypothesis} &
  \textbf{Statistical Testing and Result. } &
  \textbf{Supported?} \\ \hline
\textbf{H1}: Degree of Alignment $>$ 0 &
  \makecell[l]{
  \textbf{One-sample Student's t-test}: \\ $p<0.001$***, $d=0.521$ 
  \\ \textbf{One-sample Wilcoxon signed-rank test}: \\ $p<0.001$***, $r_{rb}=0.559$} &
  \checkmark \\
\textbf{H2}: Topic $\times$  Time $\rightarrow$ Human-AI Self-concept Alignment, the alignment is stronger in personal topic condition &
  \makecell[l]{\textbf{Repeated measures ANOVA}: \\ Interaction Effect: $p=0.011$*, $\eta_p^2=0.073$
  \\ \textbf{Bonferroni post-hoc pairwise comparisons}: \\ Personal topic: $p<0.001$***, $d=-0.244$ \\ Non-personal topic: $p=0.291$, $d=-0.087$} &
  \checkmark \\
\textbf{H3}: Post-conversation Inter-participant Self-concept Distance $<$ Baseline Inter-participant Self-concept Distance &
  \makecell[l]{\textbf{Paired samples Student's t-test}: \\ $p<0.001$***, $d=-0.305$ 
  \\ \textbf{Paired samples Wilcoxon signed-rank test}: \\ $p<0.001$***,  $r_{rb}=-0.358$} &
  \checkmark \\
\textbf{H4}: Degree of Alignment\textuparrow  $\rightarrow$ Conversation Enjoyment\textuparrow &
  \makecell[l]{\textbf{SEM}:\\ Total Effect: $p=0.028$*, $\beta=0.748$} &
  \checkmark \\
\textbf{H5}: Degree of Alignment $\rightarrow$ Perception Accuracy $\rightarrow$ Shared Reality Experience $\rightarrow$ Conversation Enjoyment &
  \makecell[l]{\textbf{SEM}: \\ Indirect Effect: $p=0.027$*, $\beta=0.188$ \\ Direct Effect: $p=0.835$, $\beta=0.051$} &
  \checkmark \\ \hline
\end{tabular}%
}
}
\end{table}

\end{document}